\documentclass{interact}

\usepackage[colorlinks=true,allcolors=blue]{hyperref}
\usepackage{setspace}
\usepackage{svg}
\usepackage[caption=false]{subfig}
\usepackage{float}
\usepackage{acro}
\usepackage{booktabs}
\usepackage{multirow}
\renewcommand{\thesection}{\Roman{section}}
\renewcommand{\thesubsection}{\Roman{section}.\Alph{subsection}}
\renewcommand{\thetable}{\Roman{table}}
\usepackage{xcolor}
\newcommand{\edit}[1]{{\color{black} #1}}
\usepackage[symbol]{footmisc}

\usepackage{algorithm}
\usepackage[noend]{algpseudocode}
\usepackage{eqparbox}
\newdimen{\algindent}
\algnewcommand\LeftComment[2]{%
\hspace{#1\algindent}$\triangleright$ \eqparbox{COMMENT}{#2} \hfill %
}

\usepackage{amsmath}
\usepackage{mathtools}
\DeclarePairedDelimiter\ceil{\lceil}{\rceil}
\DeclarePairedDelimiter\floor{\lfloor}{\rfloor}

\begin{document}

\articletype{AUTHORS' ORIGINAL MANUSCRIPT}
\title{\edit{Analysis of }Population Control Techniques for Time-Dependent and Eigenvalue Monte Carlo Neutron Transport Calculations}
\author{
\name{Ilham Variansyah\thanks{CONTACT Ilham Variansyah. Email: ijuanda@nd.edu} and Ryan G. McClarren}
\affil{University of Notre Dame, Department of Aerospace and Mechanical Engineering, Notre Dame, Indiana 46556}
}
\maketitle

\begin{abstract}
\edit{An extensive} study of population control techniques (PCTs) for time-dependent and eigenvalue Monte Carlo (MC) neutron transport calculations is presented. We define PCT as a technique that takes a censused population and returns a controlled, unbiased population. A new perspective based on an abstraction of particle census and population control is explored, paving the way to improved understanding and application of the concepts. Five distinct PCTs identified from the literature are reviewed: Simple Sampling (SS), Splitting-Roulette (SR), Combing (CO), modified Combing (COX), and Duplicate-Discard (DD). A theoretical analysis of how much uncertainty is introduced to a population by each PCT is presented. Parallel algorithms for the PCTs applicable for both time-dependent and eigenvalue MC simulations are proposed. The relative performances of the PCTs based on runtime and tally mean error or standard deviation are assessed by solving time-dependent and eigenvalue test problems. It is found that SR and CO are equally the most performant techniques, closely followed by DD.
\end{abstract}

\section{Introduction}

The Monte Carlo (MC) method is indispensable in neutron transport calculations due to its ability to perform high-fidelity, continuous-energy transport simulations with minimal approximation. MC, however, suffers from stochastic uncertainties requiring an expensive computation of a large number of neutron source samples or histories. Nevertheless, thanks to the advancement of high-performance parallel computing, the inherently parallel features of MC can be effectively exploited to a very large extent---which can significantly reduce run time to solution, particularly for the computationally expensive time-dependent neutron transport simulations \cite{cullen2016tart, sjenitzer2013dmc, leppanen2013serpent2TDMC, sweezy2014mcatkSR, russel2014geantTDMC, mylonakis2017openmcTDMC, shaukat2017mccardTDMC, molnar2019guardyan, nauchi2019tdmcWithK, ajami2021newCombing}.

During a time-dependent MC simulation, particle population size can monotonically grow or decay depending on the criticality of the system. This monotonic evolution of population makes time-dependent MC simulation particularly challenging in two different ways. First, in a supercritical system, particle population size can quickly grow beyond the limited computational resources. Additionally, some MC implementations and variance reduction techniques---such as the precursor forced decay technique in \cite{sjenitzer2013dmc} and time-dependent adaptation of the hybrid source iteration methods in \cite{willert2013hybrid, pasmann2021qmc}---may promote monotonic population growth, which raises the same issue on the limited computational memory. Second, in a subcritical system without a significant persisting external source---such as in pulsed-reactor and shut-down experiments---particle population size can quickly decay to zero, which leads to a lack of samples and yields statistically noisy tally results at later times of the simulation.

One typically uses a Population Control Technique (PCT) to address the monotonic population growth and collapse issues discussed above. PCT essentially controls the size of a particle population to be near a desired value while preserving certain statistical expectations to achieve unbiased MC simulation. In the implementation of PCT, time census is employed  to limit the population growth/collapse. The census introduces a time boundary that stops particles whenever they are about to cross it. When all particles have already hit the time boundary, the time census is completed, and PCT can be performed on the censused particles. More recent applications of PCT include the use of random particle duplication or discard \cite{leppanen2013serpent2TDMC} in Serpent 2 \cite{leppanen2015serpent2}, splitting and Russian-Roulette technique \cite{sweezy2014mcatkSR} in MCATK \cite{adams2015mcatk}, particle combing technique \cite{booth1996comb} in TRIPOLI-4 \cite{brun2015tripoli4, faucher2018tripoli4TDMC} and GUARDYAN \cite{molnar2019guardyan}, and a modified combing technique which is most recently introduced in \cite{ajami2021newCombing}.

An innovative approach to performing time-dependent MC is proposed by \cite{nauchi2019tdmcWithK}. The central idea is to re-purpose the generally available $k$-eigenvalue MC simulation infrastructure to perform time-dependent simulations. This approach works because there is a built-in population control in $k$-eigenvalue MC simulation. Besides the introduction of the $1/k$ factor on the fission operator, which is essential in achieving a steady-state configuration, simple sampling is typically performed to ensure that a certain number of particles are sampled from the fission bank and then used as the particle source for the simulation of the next fission generation. Observing the significance of that connection between the $k$-eigenvalue and time-dependent MC simulations offers an improved understanding of PCT. Such a study has been done to an extent by Cullen et al. in \cite{cullen2003aVSk}. Nevertheless, one can take advantage of this connection further by exploring potential benefits from and for both of the simulation modes.

Despite the multiple distinct PCTs proposed in the literature \cite{sjenitzer2013dmc, leppanen2013serpent2TDMC, sweezy2014mcatkSR, nauchi2019tdmcWithK, ajami2021newCombing}, documented studies in characterizing and assessing relative performances of all the identified PCTs are still very limited. A more recent effort found in \cite{sweezy2014mcatkSR} specifically compares the splitting and Russian-Roulette technique \cite{sweezy2014mcatkSR} to the particle combing technique \cite{booth1996comb}---hereafter referred to as Splitting-Roulette (SR) and Combing (CO), respectively. Sweezy et al. \cite{sweezy2014mcatkSR} propose a \emph{normalized} SR as an alternative to CO which may suffer from unwanted behavior due to possible correlations in the particle order. On the other hand, Faucher et al. \cite{faucher2018tripoli4TDMC} and Legrady et al. \cite{legrady2020pctVR} prefer the use of CO instead of SR due to the inherent bias in the normalized SR \cite{sweezy2014mcatkSR} and suggest that the unwanted behavior of CO is unlikely to occur in practice. This support for CO, or, if you will, ctenophilia, is \edit{strengthened by an assertion \cite{legrady2020pctVR} stating that the SR technique described in \cite{sweezy2014mcatkSR} is at least 2–-3 times less efficient than CO. This assertion is based on a comparative study of CO and a ``Russian roulette and splitting" technique reported in \cite{faucher2019thesis}. However, the Russian roulette and splitting technique in \cite{faucher2019thesis} seems to be different from the SR technique described in \cite{sweezy2014mcatkSR}; and for the record, the study \cite{faucher2019thesis} does not claim that the technique compared refers to \cite{sweezy2014mcatkSR}. Consistent implementation and fair comparison of the two techniques, SR \cite{sweezy2014mcatkSR} and CO \cite{booth1996comb}, would shed light on their actual relative performances.}

In this paper, we present \edit{an extensive} study on PCT. In Sec. \ref{sec:pct}, we start by making an abstraction of related concepts---i.e., particle census and population control---followed by reviewing PCTs identified from the literature. In Sec. \ref{sec:variance}, we perform an analysis to reveal the theoretical uncertainty introduced by each of the PCTs, which directly affects the performance of the technique; these theoretical uncertainties are then verified numerically. Sec. \ref{sec:parallel} presents a parallel PCT algorithm that exploits the abstraction established in Sec. \ref{sec:pct} and adapts the nearest-neighbor parallel fission bank algorithm proposed in \cite{romano2012parallelBankAlg}. In Secs. \ref{sec:test_td} and \ref{sec:test_eigen}, we implement and test the PCTs on time-dependent and eigenvalue MC neutron transport problems, respectively. Finally, Sec. \ref{sec:conclusion} summarizes the takeaways of the study. \edit{It is worth mentioning that while this paper is focused on PCT application for MC neutron transport, discussions and analyses presented in this paper are also applicable to PCT application in other transport simulations, such as the Implicit Monte Carlo thermal radiative transfer \cite{wollaber2016}}.

\section{Population Control Technique (PCT)}\label{sec:pct}

Population control can be loosely defined as any MC technique that involves altering the number of particles being simulated; this includes many variance reduction techniques (e.g., cell importance and weight window) and even the introduction of $1/k$ factor in eigenvalue simulations \cite{cullen2003aVSk}. However, in this paper, we specifically define population control as a technique that controls populations of \emph{censused} particles. In this section, we present an abstraction of particle census and population control (their definitions and how they are characterized) and then discuss distinct techniques identified from the literature.

\subsection{Particle Census}\label{sec:census}

Census is a process where we (1) stop particles, (2) remove them from the current simulation, and then (3) store them into a \emph{census bank}. Census can be performed at arbitrary steps during simulation; however, there are several triggering events that physically make sense to perform the census.

Perhaps the most obvious one is time-grid crossing. In this \emph{time census}, we stop particles whenever they are about to cross a predetermined time grid; these censused particles are then removed from the current simulation and stored into a \emph{time bank} (the census bank).

Another useful triggering event is fission emission. In this \emph{fission census}, neutrons emitted from fission reactions are removed from the current simulation and stored into a \emph{fission bank}. One can see that this is actually a standard practice that has been long used in $k$-eigenvalue MC transport simulations. We can take a step further and census not only the fission neutrons but also the scattering neutrons---this results to \emph{collision census}, which is typically used in the $c$-eigenvalue MC calculations \cite{kiedrowski2012ceig}.

There are several reasons to perform particle census. One is to limit particle population growth so that population control (discussed in more detail next) can be performed. Another reason is to allow the system (the MC model) to change---which can be geometry, composition, or parameter changes due to multi-physics feedback. Additionally, one can also see census as a manifestation of an iterative scheme performed to solve an equation---e.g., power iteration in $k$-eigenvalue problem.

It is worth noting that census time grid for population control does not necessarily need to be identical to other time grids possibly used in MC simulation. These other time grids include the one for tally scoring (also known as tally filters in some MC codes, such as OpenMC \cite{romano2015OpenMC}), time grid for variance reduction techniques (e.g., weight window and forced precursor decay \cite{sjenitzer2013dmc}), and census time grid for model change or multi-physics feedback.

\subsection{Population Control}\label{sec:popctrl}

\begin{figure}[H]
    \centering
    \includegraphics[width=0.5\textwidth]{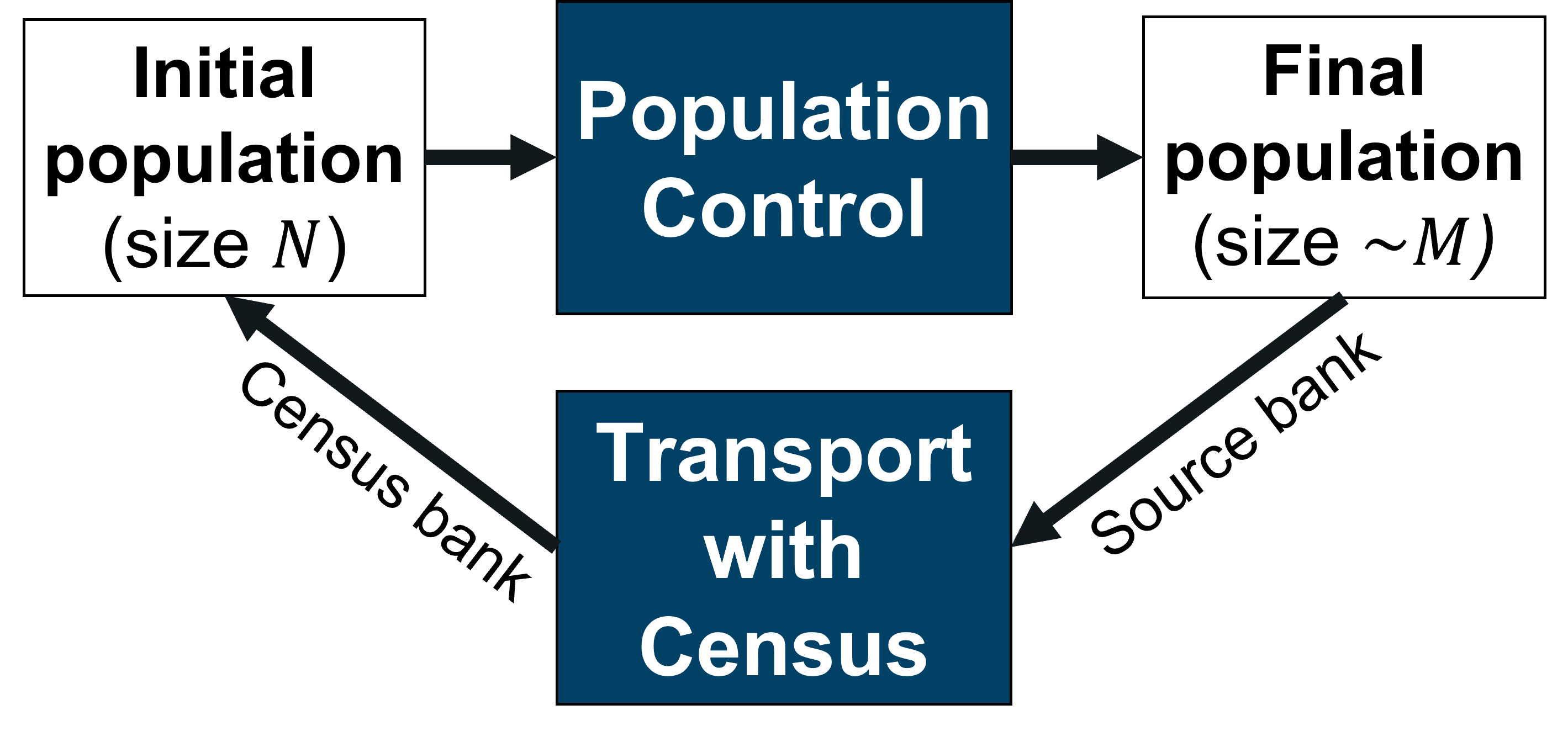}
    \caption{Illustration of census-enabled MC transport with population control.}
    \label{fig:popctrl}
\end{figure}

Given an \emph{initial population} of size $N$, the objective of population control is to return a controlled \emph{final population} with a size around, or exactly at, a predetermined value $M$, as illustrated in Fig. \ref{fig:popctrl}. In a supercritical system, typically $N>M$; while in a subcritical one, $N<M$. The final population is then used as the source bank for the successive \emph{census-enabled} transport simulation, during which a census bank is populated by a certain census mechanism (e.g., time census or fission census, as discussed in Sec. \ref{sec:census}). Once the transport is completed (i.e., both source and secondary particle banks are exhausted), the census bank becomes the initial population to be controlled by a PCT of choice. It is evident that population control does not care about what kind of transport simulation is being performed, whether it is a time-dependent fixed-source or an eigenvalue one. This also implies that any PCT can be used in any kind of transport simulation; as a particular example, one can use the particle combing technique \cite{booth1996comb} in $k$-eigenvalue simulation.

The final population basically consists of copies of the initial particles, but how many times a particle gets copied will differ between particles, and some particles may not get copied at all. The procedure on determining how many times each initial particle get copied to the final population is the essence of PCT and has to be done in a way such that the MC simulation is not biased---i.e., the expectations of the population actions, and thus the expectations of simulation tally results, are preserved.

The only requirement for a PCT to be unbiased is to preserve the expected weight of each particle in the initial population. That is, for initial particle $i$ having weight $w_i$:
\begin{equation}\label{eq:unbiasReq}
  \boldsymbol{E}[C_i] = w_i,\qquad i=1, 2, ..., N,
\end{equation}
\begin{equation}
  C_i = d_i w_i',
\end{equation}
where $\boldsymbol{E}[\cdot]$ denotes the expectation of a random variable argument, $d_i$ is the number of copies of particle $i$ in the final population, $w_i'$ is the controlled weight assigned to the copies of particle $i$, and $C_i$ is the total weight represented by the copies of particle $i$ in the final population.

Now that we have described the minimum requirements---i.e., controlling population size from $N$ to around $M$, while ensuring that Eq. \eqref{eq:unbiasReq} holds---we next point out two desirable characteristics of PCT.

The first is that we wish to have a low uncertainty in $C_i$. \edit{In particular, we want its standard deviation, $\boldsymbol{\sigma}[C_i]$, to be as low as possible.} In the absence of PCT, each initial particle $i$ is copied once to the final population, and we will have $C_i=w_i$ and $\boldsymbol{\sigma}[C_i]=0$; however, if a PCT is being used, $\boldsymbol{\sigma}[C_i]\geq0$. \edit{The value of $\boldsymbol{\sigma}[C_i]$ affects the variance of the actions of particle $i$ in the MC simulation, and later we numerically demonstrate that it ultimately affects the simulation tally results.}

The second desirable characteristic is that we would like our PCT to preserve the initial population total weight $W$ as much as possible; in other words, if $W'$ is the final population total weight:
\begin{equation}\label{eq:W}
  W = \sum_{i=1}^{N}w_i,
\end{equation}
\begin{equation}\label{eq:Wp}
  W' = \sum_{i=1}^{N}C_i,
\end{equation}
and we would like $W'$ to be close or equal to $W$. Booth \cite{booth1996comb} suggests that such strict equality of $W'=W$ is generally unimportant for neutron and photon transport, but it may be very important in charged particle transport. Therefore, we consider it a desirable characteristic, not a requirement, of PCT.

As a remark, PCT is a technique that takes an initial population of size $N$ and total weight $W$ and returns a controlled final population that:
\begin{enumerate}
    \item has a size equal or close to $M$,
    \item preserves the expected total weight of each  particle (i.e., satisfies Eq. \eqref{eq:unbiasReq}, $\boldsymbol{E}[C_i]=w_i$),
    \item has a low $\boldsymbol{\sigma}[C_i]$, and
    \item has a total weight equal or close to $W$.
\end{enumerate}
We note that Point (1) is the objective of PCT, Point (2) is the requirement for unbiased PCT, and Points (3) and (4) are desirable characteristics.

\subsection{The PCTs}\label{sec:pcts}

Per our literature study, we identify five distinct PCTs: (1) Simple Sampling (SS), (2) Duplicate-Discard (DD) \cite{leppanen2013serpent2TDMC}, (3) Splitting-Roulette (SR) \cite{sweezy2014mcatkSR}, (4) Particle Combing (CO) \cite{booth1996comb}, and (5) Modified Particle Combing (COX) \cite{ajami2021newCombing}. Additionally, there are three different sampling bases with which each of the PCTs can be implemented: uniform, weight-based, and importance-based sampling.

\subsubsection{Combing (CO)}\label{sec:CO}

\begin{figure}[H]
    \centering
    \includegraphics[width=1.0\textwidth]{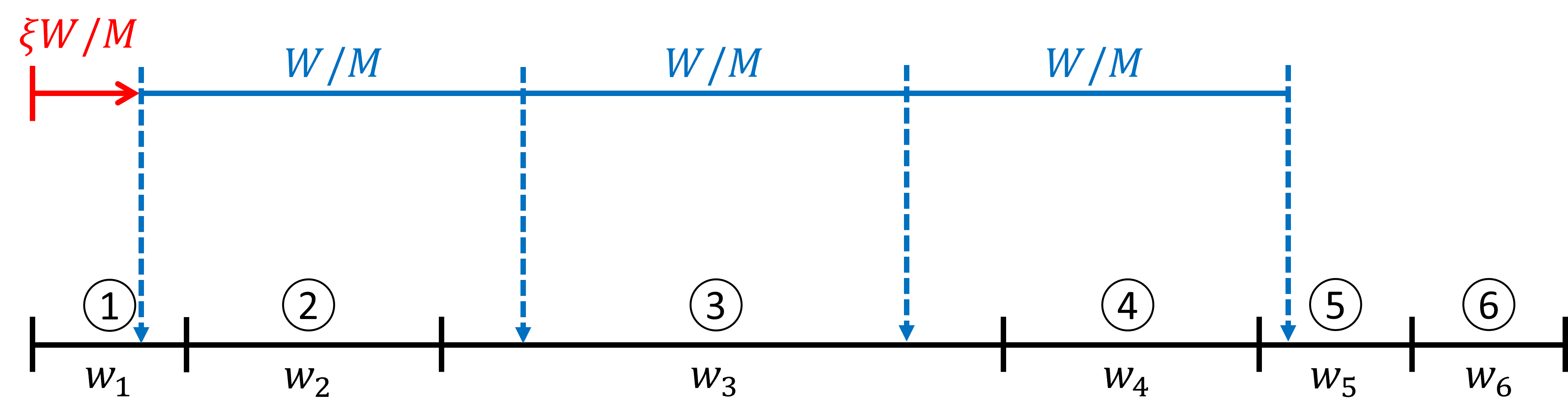}
    \caption{Weight-based CO with initial and final population size $N$=6 and $M$=4 \edit{(adapted from Fig. 1 in \cite{booth1996comb})}.}
    \label{fig:COW}
\end{figure}

Perhaps the most standardized\footnote[1]{\edit{CO is the only technique that has a single, well-known origin to refer to (\cite{booth1996comb}), a proper name, and a clear, unambiguous procedure.}} PCT is the particle combing technique (CO). Per our classification, the “Simple Comb” proposed by Booth (Section II in \cite{booth1996comb}) is weight-based CO. CO techniques are best explained with graphical illustrations. Let us consider a population control problem with $N$=6 and $M$=4. Weight-based CO combs the population as shown in Fig. \ref{fig:COW}, where $\xi$ is a random number (from 0 to 1) used to determine the offset of the initial tooth of the comb. Once the initial tooth location is set, the remaining teeth are spaced $W/M$ apart. Per Fig. \ref{fig:COW}, Particles 1 and 5 are copied once, Particle 3 is copied twice, and Particles 2, 4, and 6 are not copied at all. To ensure unbiased MC simulation (c.f.\ Eq.~\eqref{eq:unbiasReq}), the copies of particle $i$ are assigned with weight $w_i'=W/M$. 

Booth also proposes the “Importance-Weighted Comb” (Section III in \cite{booth1996comb}), which per our classification is importance-based CO. Importance-based CO is similar to the weight-based CO shown in Fig. \ref{fig:COW}, but instead of using $w_i$ for the particle axis, $W/M$ for the distance between teeth, $\xi W/M$ for the offset of the comb, and final weight $w_i'=W/M$, we respectively use $u_i$, $U/M$, $\xi U/M$, and $w_i'=U/(MI_i)$---where $u_i=I_iw_i$ is the product of importance $I_i$ and weight of particle $i$, and $U=\sum_{i}u_i$ is the total of the product.

\begin{figure}[H]
    \centering
    \includegraphics[width=1.0\textwidth]{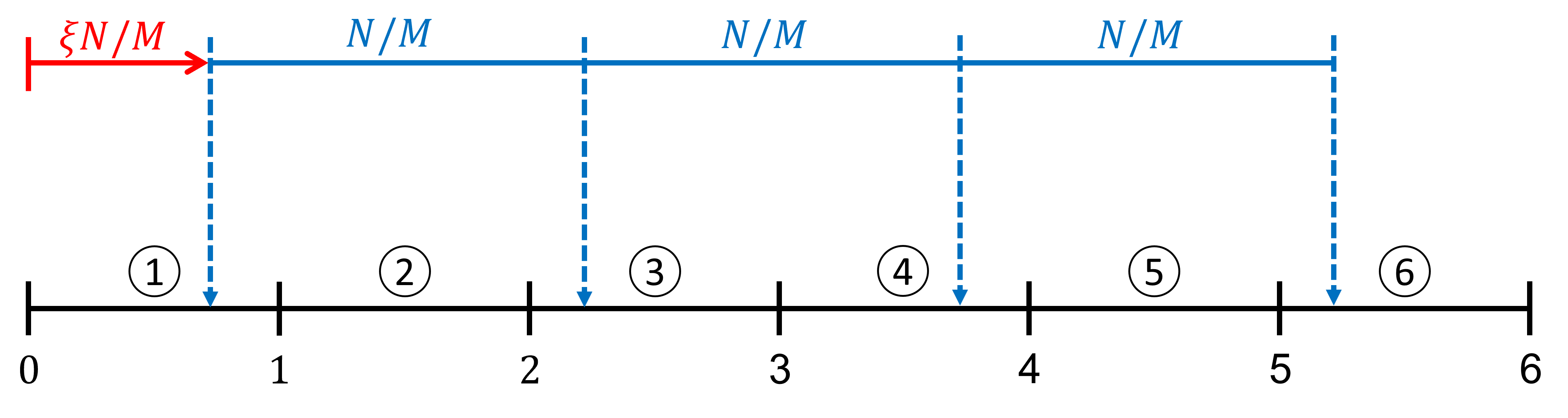}
    \caption{Uniform CO with initial and final population size $N$=6 and $M$=4.}
    \label{fig:COU}
\end{figure}

Now we discuss the other variant of CO: uniform CO. \edit{Uniform CO treats particles equally, regardless of their weight or importance.} Uniform CO combs the initial particles as shown in Fig. \ref{fig:COU}. Per Fig. \ref{fig:COU}, each of Particles 1, 3, 4, and 6 is copied once, while Particles 2 and 5 are not copied at all. To ensure unbiased MC simulation (Eq.~\eqref{eq:unbiasReq}), copies of particle $i$ are assigned with weight $w_i'=(N/M)w_i$. We believe that this uniform variant of CO (as well as those of other PCTs) has never been articulated in the literature. A discussion on the significance of the PCT sampling bases (uniform, weight-based, or importance-based) is given later in Sec. \ref{sec:samplingbasis}.

\subsubsection{Modified Combing (COX)}\label{sec:COX}

\begin{figure}[H]
    \centering
    \includegraphics[width=1.0\textwidth]{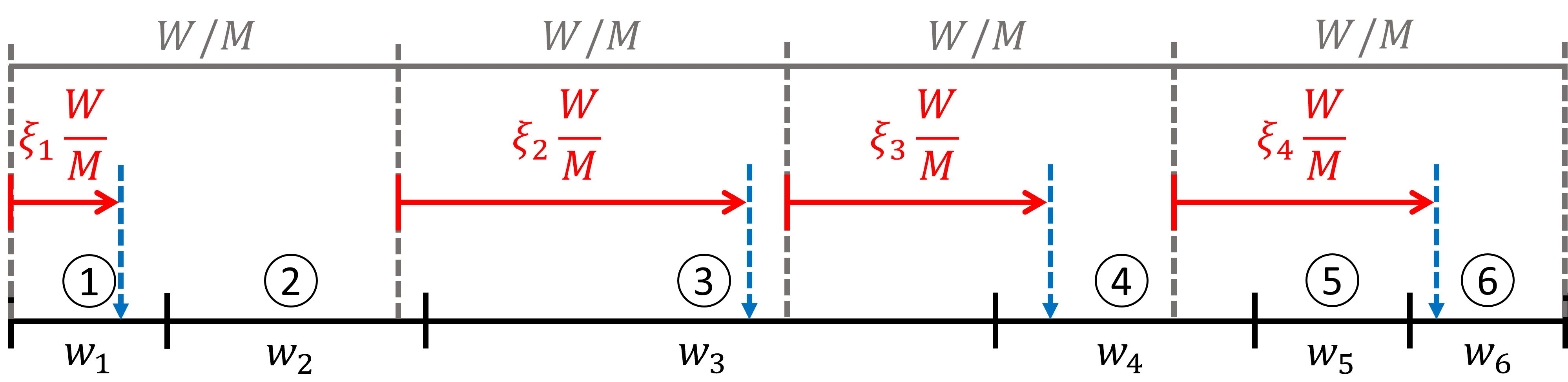}
    \caption{Weight-based COX with initial and final population size $N$=6 and $M$=4.}
    \label{fig:COXW}
\end{figure}

A modification of CO is recently proposed by Ajami et al. \cite{ajami2021newCombing}. Different from the weight-based CO shown in Fig. \ref{fig:COW}, the weight-based COX combs the initial particle as shown in Fig. \ref{fig:COXW}. In COX, instead of having uniformly-spaced teeth and sampling the offset of the whole comb, we allow the teeth to be non-uniformly spaced by offsetting each tooth with a different random number. The controlled weight $w_i'$ assigned to the particle copies to ensure unbiased MC simulation (Eq. \eqref{eq:unbiasReq}) are identical to those of CO.

Ajami et al. \cite{ajami2021newCombing} provides limited discussion and demonstration on how COX compares to CO. In Sec. \ref{sec:correlation}, we discuss how COX may actually avoid a \edit{particular} drawback of CO; and then in Sec. \ref{sec:variance}, we discuss how that remedy comes at a significant expense.

\subsubsection{Splitting-Roulette (SR)}\label{sec:SR}

Sweezy et al. \cite{sweezy2014mcatkSR} proposes the weight-based splitting-roulette (SR). In SR, we assign each initial particle $i$ with \edit{\emph{splitting number} $s_i$}. For uniform, weight-based, and importance-based SR, the values for $s_i$ are respectively $M/N$, $w_i/(W/M)$, and $u_i/(U/M)$. We split each particle $i$ into $\floor*{s_i}+1$ copies, and then Russian-roulette the last copy with \edit{surviving probability} $s_i-\floor*{s_i}$; the function $\floor*{\cdot}$ denotes the floor function, which produces the greatest integer not greater than the variable argument. Finally, to ensure unbiased MC simulation, the surviving particle copies are assigned with controlled weight $w_i'$, which happen to be identical to those of CO techniques.

SR techniques neither exactly produce a final population of size $M$ nor exactly preserve the initial total weight $W$---however, they preserve the expectations. To exactly preserve the population's total weight $W$, Sweezy et al. suggest performing a weight normalization at the end of SR. This weight normalization can be applied to other PCTs that do not exactly preserve the population’s total weight as well (e.g., uniform and importance-based CO). The significance of this PCT weight normalization is further discussed later in Sec. \ref{sec:normalization}.

\subsubsection{Simple Sampling (SS)}\label{sec:SS}

Simple sampling (SS) is the typical PCT employed in $k$-eigenvalue MC simulations \cite{romano2012parallelBankAlg}. In SS, we simply sample $M$ particles from the initial population to be the final population. For uniform SS, all particles have a uniform probability to be sampled at each draw; while for weight-based and importance-based SS, the probability for a particle to be sampled at each draw is proportional to its weight $w_i$ and the product of its weight and importance $u_i$, respectively. Finally, to ensure unbiased MC simulation, the sampled particles are assigned with controlled weight $w_i'$ which values happen to be identical to those of the other PCTs.

\subsubsection{Duplicate-Discard (DD)}\label{sec:DD}

We identify the PCT proposed by Leppänen in \cite{leppanen2013serpent2TDMC} as the uniform duplicate-discard technique (DD), due to its mechanism of randomly duplicating or discarding particles to achieve the desired population size. \edit{In particle duplication (for $N<M$), we first copy each initial particle once to the final population, and then, on top of that, randomly sample $M-N$ particles from the initial population to be copied to the final population. In particle discard (for $N>M$), we randomly sample $N-M$ particles from the initial population; the sampled particles do not get copied to the final population, while the rest are copied once.} Finally, the controlled weight $w_i'$ that satisfies the unbiased MC simulation requirement is identical to that of the other uniform PCTs: $(N/M)w_i$.

One can improve the particle duplication of the uniform DD. Instead of keeping a copy of the initial population and then sampling $M-N$ additional particles, we keep $\floor{M/N}$ copies and sample only $(M\bmod N)$ particles (we note that ``$\bmod$'' denotes the remainder operator, such that $(M\bmod N)=M-\floor{M/N}N$). This improvement reduces both the number of samplings performed and the uncertainty introduced by the PCT.

\subsection{Additional Notes on the PCTs}\label{sec:pct_notes}

\subsubsection{PCT Sampling Basis}\label{sec:samplingbasis}

As mentioned earlier, each of the five distinct PCTs (CO, COX, SR, SS, and DD) can be implemented with three different sampling bases: uniform, weight-based, and importance-based sampling.

The computational procedures of the uniform sampling PCTs are the simplest, followed by their respective weight-based and then importance-based counterparts. As an example, uniform CO (Fig. \ref{fig:COU}) is simpler than the weight-based CO (Fig. \ref{fig:COW}) as it does not require some binary search to determine where exactly each tooth falls.

If the initial population has a uniform weight, the weight-based sampling is identical to the uniform sampling, since $W=Nw_i$. However, if the initial particles have varying weights, the weight-based sampling simultaneously functions as a variance reduction technique as well: particles having relatively large weights tend to be split into multiple copies, which leads to variance reduction; on the other hand, particles with relatively low weights tend to be Russian-rouletted, which may lead to more efficient computation by not spending time tracking small-weight particles. Nevertheless, particle weight does not necessarily indicate particle importance. If the initial particles are assigned with some importance values, the importance-based sampling offers more effective variance reduction than the weight-based.

One may argue that uniform sampling is the least optimal as it assigns all particles an identical splitting number or surviving probability regardless of their weights and importance. However, uniform sampling can be the most optimum choice in two cases. The first is when the population has a uniform weight and unknown importance, which is the case in a fixed-source problem without any variance reduction technique and in the typical $k$-eigenvalue simulation where all the fission neutrons are emitted with uniform weight. The second case is when the MC simulation is already equipped with some variance reduction techniques, such as the weight window or the uniform fission site method \cite{hunter2013ufs}, because particle distribution and weight profile of the population would be optimized already, such that particles can be treated equally by the PCT. In other words, avoiding redundancy in variance reduction effort. In particular, in the application of an effective weight window or the uniform fission site method, the use of weight-based sampling may actually ruin the already optimized particle distribution.

The interplay between PCT and variance reduction technique briefly described above is outside the scope of this study. Furthermore, while the theoretical analysis performed in Sec. \ref{sec:variance} is applicable to all sampling bases, only the uniform PCTs are implemented and tested in Secs. \ref{sec:parallel}--\ref{sec:test_eigen}.

\subsubsection{Correlation Issue in CO}\label{sec:correlation}

In Sec. \ref{sec:pcts}, it is interesting to observe that CO techniques only require one random number to perform the population control (as a comparison, SS and SR respectively require $M$ and $N$ random numbers); in other words, a single random number determines the fate of all particles in the population. This unfortunately yields correlation in the particle sampling. As an example, Particles 1 and 2 in Fig. \ref{fig:COW} will never be sampled together. This correlation may produce unwanted behavior depending on how the initial particles are ordered.

\begin{figure}[H]
    \centering
    \includegraphics[width=0.75\textwidth]{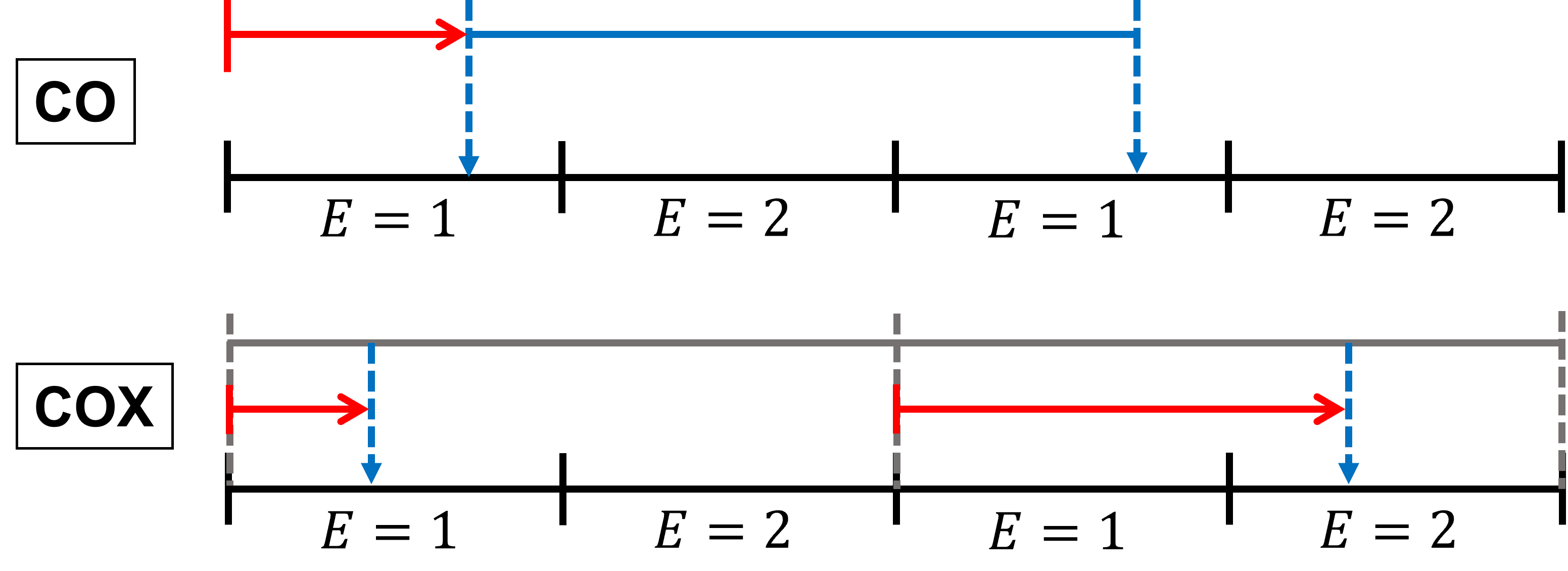}
    \caption{An illustration of the correlation issue in CO and how it is remedied in COX; \edit{$E$ indicates particle's energy in MeV.}}
    \label{fig:CO_issue}
\end{figure}

Sweezy et al. \cite{sweezy2014mcatkSR} provide an illustrative demonstration of such possible unwanted behavior in CO, which is shown in the upper part of Fig. \ref{fig:CO_issue}. In this postulated PCT problem, we wish to select 2 particles from an ordered initial population of size 4. The initial population consists of alternating 1-MeV and 2-MeV particles, all of which have uniform weight. If we apply CO, we will have a final population with either all 1-MeV or all 2-MeV particles. However, this behavior does not necessarily make the MC simulation biased, because each initial particle is still treated fairly individually—i.e., Eq. \eqref{eq:unbiasReq} is still satisfied. If one were to run the simulation in multiple batches---which is necessary to get a measure of result uncertainty in a census-enabled MC simulation---we would be running half of the batches with all 1-MeV particles and 2-MeV on the other half. While such behavior may result in a larger tally variance, the expectation is still preserved. Outside this postulated PCT problem, some extent of physics is naturally embedded in the particle population order (e.g., adjacent particles may be originated from the same emission event). However, there has never been any observable effect of this correlation issue in the practical application of CO \cite{sjenitzer2013dmc, faucher2018tripoli4TDMC, faucher2019thesis, molnar2019guardyan, legrady2020pctVR}.

If one wishes to eliminate this possible correlation issue, the initial population order must be randomized before CO is applied. However, in massively parallel computation with reproducibility requirement, this pre-randomization process will require a large number of communications, which may ruin the parallel scalability of the simulation.

The modified combing technique COX proposed by Ajami et al. \cite{ajami2021newCombing}, to some extent, remedies this correlation issue as demonstrated in the lower part of Fig. \ref{fig:CO_issue}. Nevertheless, this remedy comes at the expense of increasing $\boldsymbol{\sigma}[C_i]$, which is discussed later in Sec. \ref{sec:variance}.

\subsubsection{PCT Weight Normalization}\label{sec:normalization}

Some PCTs---i.e, uniform and importance-based PCTs, and all SR techniques---do not exactly preserve the population total weight $W$. However, the expectation of the total weight is still preserved because
\begin{equation}
  \boldsymbol{E}[W'] = \sum_{i=1}^{N}\boldsymbol{E}[C_i] = W,
\end{equation}
where the first and the second equalities respectively use Eqs. \eqref{eq:Wp} and \eqref{eq:W}. 

To exactly preserve $W$, Sweezy et al. \cite{sweezy2014mcatkSR} suggest performing weight normalization after population control is performed. This is done by multiplying all of the final particles with the factor $W/W'$, so that $C_i^\text{(norm.)}=(W/W')C_i$. Unfortunately, this PCT weight normalization introduces bias as Eq. \eqref{eq:unbiasReq} is now violated:
\begin{equation}
  \boldsymbol{E}\left[C_i^\text{(norm.)}\right] = \boldsymbol{E}\left[\frac{W}{W'}C_i\right] = \boldsymbol{E}\left[\frac{W}{W'}\right]w_i \geq w_i,
\end{equation}
where the inequality comes from Jensen's inequality \cite{jensen1960, sweezy2014mcatkSR} suggesting $\boldsymbol{E}[W/W']\geq1$. Nevertheless, it can be seen that by using a large number of particles, the bias in the normalized PCTs can be minimized; however, it is also the case for the lack of exact total weight preservation in the non-normalized PCTs. In other words, PCT weight normalization suggested in \cite{sweezy2014mcatkSR} is only recommended if preserved total weight is more important than unbiased MC simulation.

\subsubsection{More Advanced PCTs}

The techniques considered in this work are those of basic PCTs. More advanced PCTs include the one proposed by Booth in Section IV of \cite{booth1996comb}, which introduces the idea of partial population weight adjustment---an unbiased alternative to the weight normalization proposed by Sweezy et al. \cite{sweezy2014mcatkSR} (see Sec. \ref{sec:normalization})---to exactly preserve the population’s total weight $W$. This partial adjustment is technically more advanced than the weight normalization technique; it introduces tunable parameters (i.e., the adjusted partial population size and the number of recursive partial adjustments) and additional challenges for parallel computing implementation. While the proposed partial population weight adjustment is applied to the importance-based CO in \cite{booth1996comb}, it basically can be applied to other PCTs that do not exactly preserve $W$ as well. Other developments of advanced PCTs include the more recent study by Legrady et al. \cite{legrady2020pctVR}, which introduces several advanced CO techniques specifically improved for extensive variance reduction.

\section{Uncertainty Introduced by PCT}\label{sec:variance}

\subsection{Theoretical Analysis}\label{sec:sigmar_theory}

By determining the first and second moments of $C_i$ (the total weight of the copies of initial particle $i$ in the final population), we can determine the variance introduced by a PCT:
\begin{equation}\label{eq:var}
  \boldsymbol{Var}[C_i] = \boldsymbol{E}[C_i^2] - \boldsymbol{E}[C_i]^2.
\end{equation}
Another and perhaps more illustrative quantity is the \emph{relative} uncertainty (standard deviation) introduced by the PCT to each particle $i$ in the initial population:
\begin{equation}\label{eq:sigmar}
  \boldsymbol{\sigma_r}[C_i] = \frac{\boldsymbol{\sigma}[C_i]}{w_i} = \frac{1}{w_i}\sqrt{\boldsymbol{Var}[C_i]}.
\end{equation}

Unless normalized (as discussed in Sec. \ref{sec:normalization}), all of the identified PCTs (SS, SR, CO, COX, and DD) are unbiased, which means $\boldsymbol{E}[C_i]=w_i$. However, the second moments $\boldsymbol{E}[C_i^2]$ of the PCTs may be different and thus become the key to determine the relative performance on how large uncertainty $\boldsymbol{\sigma_r}[C_i]$ is introduced by the techniques.

In SR (described in Sec. \ref{sec:SR}), each initial particle $i$ is either copied $\floor{s_i}+1$ times with a probability of $s_i-\floor{s_i}$, or otherwise copied $\floor{s_i}$ times. This suggests
\begin{equation}
  \boldsymbol{E}[C_i^2]_{\text{SR}} = \left(s_i-\floor{s_i}\right)\left[(\floor{s_i}+1)w_i'\right]^2 + \left[1-(s_i-\floor{s_i})\right]\left(\floor{s_i}w_i'\right)^2,
\end{equation}
\begin{equation}\label{eq:sigmar_SR}
  \boldsymbol{\sigma_r}[C_i]_{\text{SR}} = \frac{1}{s_i} \sqrt{-s_i^2 + (2\floor{s_i}+1)s_i - (\floor{s_i}^2+\floor{s_i})},
\end{equation}
where we note that $w_i'=w_i/s_i$.

In CO (described in Sec. \ref{sec:CO}), each initial particle $i$ is either copied $\ceil{s_i}-1$ times with a probability of $\ceil{s_i}-s_i$, or otherwise copied $\ceil{s_i}$ times. The quantity $s_i$ (splitting number) used in this context happens to be identical to that of SR; and the function $\ceil*{\cdot}$ denotes the ceiling function, which produces the smallest integer not smaller than the variable argument. Following the similar process to that of SR in the previous paragraph, we obtain
\begin{equation}\label{eq:sigmar_CO}
  \boldsymbol{\sigma_r}[C_i]_{\text{CO}} = \frac{1}{s_i} \sqrt{-s_i^2 + (2\ceil{s_i}-1)s_i - (\ceil{s_i}^2-\ceil{s_i})}.
\end{equation}

In SS (described in Sec. \ref{sec:SS}), each particle $i$ can be copied multiple times up to $M$; this means
\begin{equation}
  \boldsymbol{E}[C_i^2]_{\text{SS}} = \sum_{j=0}^{M} \left(\genfrac{}{}{0pt}{0}{M}{j}\right) \left(\frac{s_i}{M}\right)^j \left(1-\frac{s_i}{M}\right)^{M-j} \left(jw_i'\right)^2,
\end{equation}
where we use the same definition of $s_i$ used in the other PCTs. Per binomial theorem, we can find that
\begin{equation}
  \boldsymbol{Var}[C_i]_{\text{SS}} = s_i\left(1-\frac{s_i}{M}\right){w_i'}^2,
\end{equation}
and thus
\begin{equation}\label{eq:sigmar_SS}
  \boldsymbol{\sigma_r}[C_i]_{\text{SS}} = \sqrt{ \left(\frac{1}{s_i}-\frac{1}{M}\right)}
  \approx \sqrt{\frac{1}{s_i}},
\end{equation}
where the approximation is due to the fact that typically $s_i\ll M$ (or equivalently $N\gg1$ for uniform PCTs).

In uniform DD (described in Sec. \ref{sec:DD}), we have two different cases. In the case of $N>M$, we uniformly discard $N-M$ particles from the initial population. Therefore, particle $i$ has to survive all of the discard draws to get copied once, otherwise it will not get copied at all. This means, for $N>M$ we have
\begin{equation}\label{eq:DDUsup}
  \boldsymbol{E}[C_i^2]_{\text{DD}} = \left(\frac{N-1}{N}\times\frac{N-2}{N-1}\times...\times\frac{M}{M+1}\right){w_i'}^2 = \frac{M}{N}{w_i'}^2,
\end{equation}
\begin{equation} \label{eq:sigmar_DDU_super}
  \boldsymbol{\sigma_r}[C_i]_{\text{DD}} = \sqrt{\frac{1}{s_i}-1},
\end{equation}
where again we use the same definition of $s_i$ used in the other PCTs. On the other hand, in the case of $N<M$, DD keeps $\left\lfloor M/N \right\rfloor$ copies of the initial population, and then uniformly draw a particle duplicate $(M \bmod N)$ times out of it. This process is similar to that of SS, except that we sample $(M \bmod N)$ particles instead of $M$ particles and we pre-keep $\left\lfloor M/N \right\rfloor$ copies of each initial particle. This gives
\begin{equation} \label{eq:sigmar_DDU_sub}
  \boldsymbol{\sigma_r}[C_i]_{\text{DD}} \approx \sqrt{\left(1-\frac{1}{s_i}\floor{s_i}\right)\frac{1}{s_i}},
\end{equation}
where the approximation is again due to $N\gg1$.

In COX (described in Sec. \ref{sec:COX}), things are more involved in that deriving the relative uncertainty $\boldsymbol{\sigma_r}[C_i]$ is not as straightforward. First, let us observe how Fig. \ref{fig:COXW} of COX differs from Fig. \ref{fig:COW} of CO. We can see that Particle 1 suffers from the same uncertainty in both methods; in this case, both CO and COX introduce identical uncertainty to Particle 1. However, it is not the case for the other particles. For example, in CO, Particle 2 has only two possibilities, get copied once or not at all; but in COX, Particle 2 has an additional possibility, which is to get copied twice. Due to this additional possibility, COX introduces higher uncertainty to Particle 2 than CO does. Similar findings can be observed for Particle 3. These observations indicate that $\boldsymbol{\sigma_r}[C_i]_{\text{COX}} \geq \boldsymbol{\sigma_r}[C_i]_{\text{CO}}$, depending on how the particle $i$ is located relative to the comb grid (the broken line in Fig. \ref{fig:COXW}).

\begin{figure}[H]
    \centering
    \includegraphics[width=0.5\textwidth]{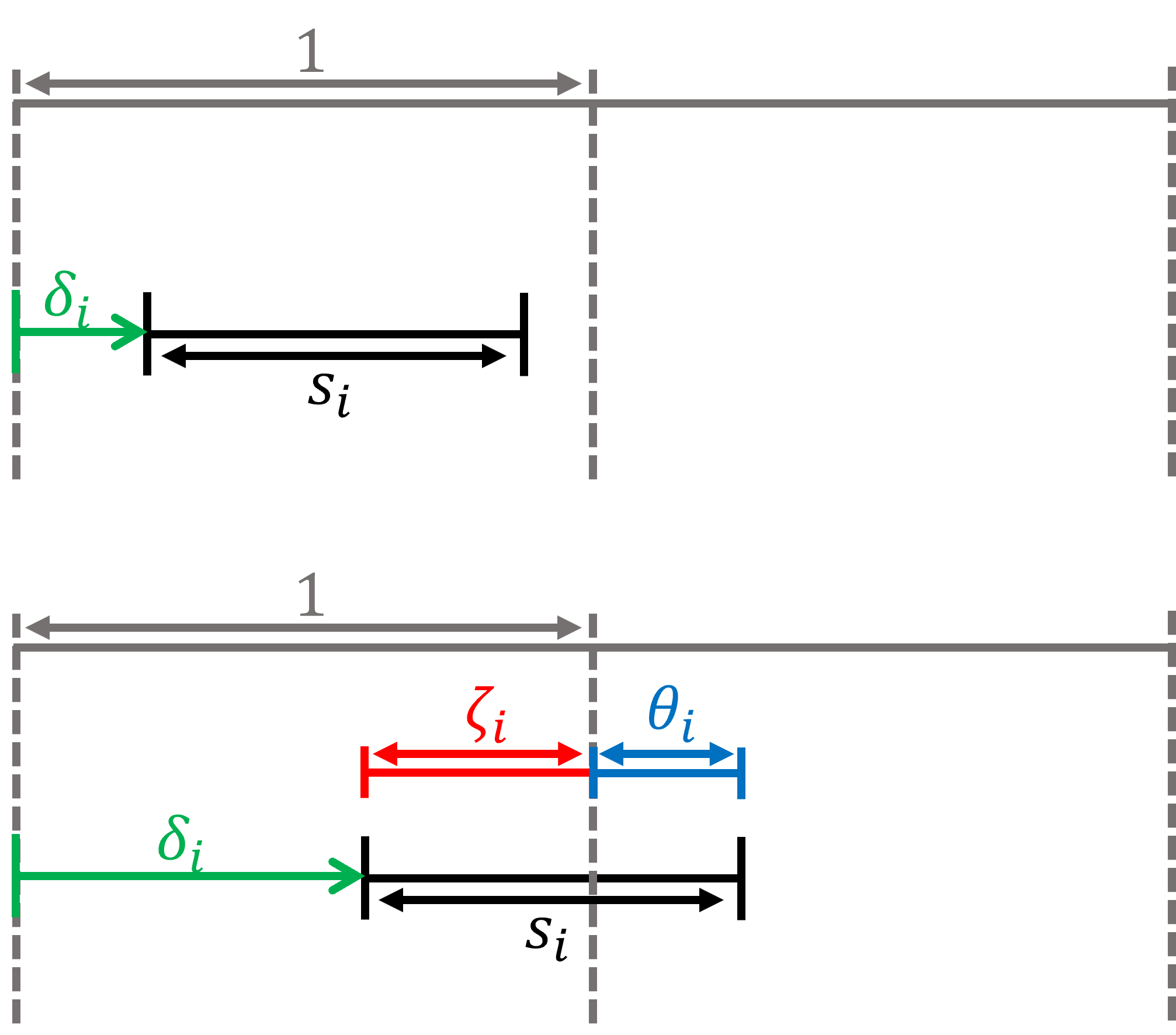}
    \caption{Illustration of particles with $s_i\leq1$ located $\delta_i$ away from COX comb grid (the broken lines).}
    \label{fig:cox_sigmar1}
\end{figure}

\begin{figure}[H]
    \centering
    \includegraphics[width=0.75\textwidth]{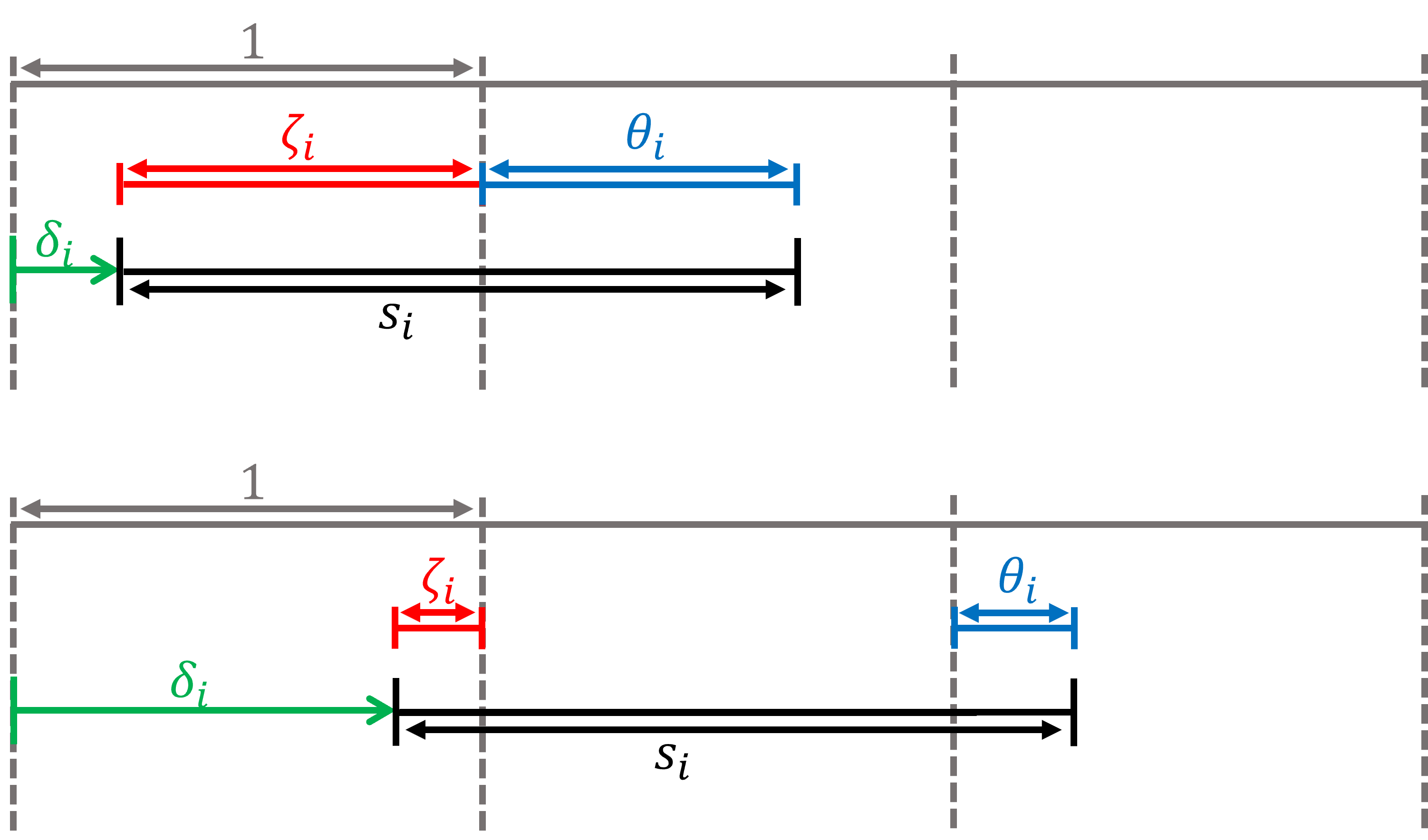}
    \caption{Illustration of particles with $s_i\geq1$ located $\delta_i$ away from COX comb grid (the broken lines).}
    \label{fig:cox_sigmar2}
\end{figure}

Figures. \ref{fig:cox_sigmar1} and \ref{fig:cox_sigmar2} illustrate different situations of how particles can be located relative to the COX comb grid. The Particle 2 and Particle 3 cases discussed in the previous paragraph are illustrated by the lower parts of Figs. \ref{fig:cox_sigmar1} and \ref{fig:cox_sigmar2}, respectively. We note that we use a unit-spaced comb grid and the same definition of $s_i$ used in other PCTs; this makes the analysis applicable for COX with any sampling basis. Symbols on the figures---i.e., $\zeta_i=1-\delta_i$ and $\theta_i=s_i+\delta_i-\ceil{s_i}$---serve as key quantities to derive $\boldsymbol{E}[C_i^2]_{\text{COX}}$ as a function of the comb offset $\delta_i$. By observing the figures, we found that $\boldsymbol{E}[C_i^2]_{\text{COX}}$ (and thus $\boldsymbol{\sigma_r}[C_i]_\text{COX}$) is dependent on $\delta_i$, and the dependency is periodic with a unit period in $\delta_i$.

\begin{figure}[H]
    \centering
    \includegraphics[width=0.5\textwidth]{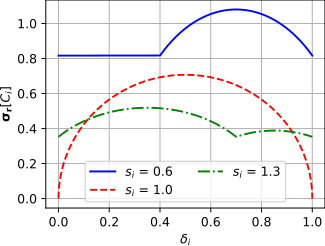}
    \caption{Theoretical relative uncertainty $\boldsymbol{\sigma_r}[C_i]$ of COX as a function of $\delta_i$ at different values of $s_i$.}
    \label{fig:cox_sigmar}
\end{figure}

On the upper part of Fig. \ref{fig:cox_sigmar1}, we have $s_i\leq1$ and $0\leq\delta_i\leq1-s_i$; in this case, COX and CO are identical. On the lower part of Fig. \ref{fig:cox_sigmar1}, we have $s_i\leq1$ and $1-s_i<\delta_i\leq1$; in this case, we have
\begin{equation}
  \boldsymbol{E}[C_i^2]_{\text{COX}} = 
  \zeta_i\theta_i (2w_i')^2 +
  (\zeta_i+\theta_i-2\zeta_i\theta_i)(w_i')^2.
\end{equation}
On the upper part of Fig. \ref{fig:cox_sigmar2}, we have $s_i\geq1$ and $0<\delta_i\leq\ceil{s_i}-s_i$; in this case, we have
\begin{multline}
  \boldsymbol{E}[C_i^2]_{\text{COX}} = 
  \zeta_i\theta_i (\ceil{s_i}w_i')^2 +
  (\zeta_i+\theta_i-2\zeta_i\theta_i)\left[\left(\ceil{s_i}-1\right)w_i'\right]^2 \\+
  (1-\zeta_i)(1-\theta_i)\left[\left(\ceil{s_i}-2\right)w_i'\right]^2.
\end{multline}
Finally, on the lower part of Fig. \ref{fig:cox_sigmar2}, we have $s_i\geq1$ and $\ceil{s_i}-s_i<\delta_i\leq1$; in this case, we have
\begin{multline}
  \boldsymbol{E}[C_i^2]_{\text{COX}} = 
  \zeta_i\theta_i \left[\left(\ceil{s_i}+1\right)w_i'\right]^2 +
  (\zeta_i+\theta_i-2\zeta_i\theta_i)(\ceil{s_i}w_i')^2 \\+
  (1-\zeta_i)(1-\theta_i)\left[\left(\ceil{s_i}-1\right)w_i'\right]^2.
\end{multline}
Fig. \ref{fig:cox_sigmar} shows the resulting $\boldsymbol{\sigma_r}[C_i]$ of COX as a function of $\delta_i$ at different values of $s_i$.

\begin{figure}[H]
    \centering
    \includegraphics[width=0.75\textwidth]{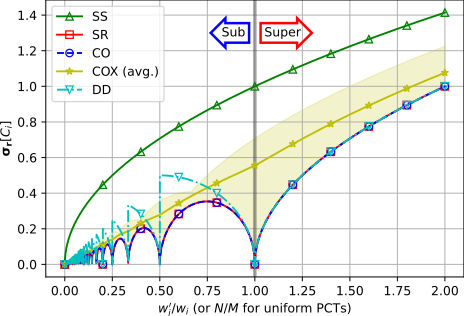}
    \caption{Theoretical relative uncertainty $\boldsymbol{\sigma_r}[C_i]$ introduced by different PCTs.}
    \label{fig:sigmar}
\end{figure}

The derived theoretical relative uncertainty $\boldsymbol{\sigma_r}[C_i]$ of the PCTs---i.e., Eq. \eqref{eq:sigmar_SR} for SR, Eq. \eqref{eq:sigmar_CO} for CO, Eq. \eqref{eq:sigmar_SS} for SS, and Eqs. \eqref{eq:sigmar_DDU_super} and \eqref{eq:sigmar_DDU_sub} for DD---are plotted in Fig. \ref{fig:sigmar}. Different to those of the other PCTs, $\boldsymbol{\sigma_r}[C_i]$ of COX is dependent on $\delta_i$ as shown in Fig. \ref{fig:cox_sigmar}; thus, in Fig. \ref{fig:sigmar}, we plot its average value and shade the region (min to max) of its possible values. The x-axis is chosen to be $1/s_i$, which is equivalent to the ratio $w_i'/w_i$---or $N/M$ for the uniform PCTs. This x-axis effectively represents a measure of the system's population growth, which is dependent on the system criticality and the census frequency. Roughly speaking, one can say that $N/M$ increases with the criticality of the system as illustrated with the arrows in the figure.

The larger $\boldsymbol{\sigma_r}[C_i]$, the larger the uncertainty introduced by the PCTs, which may lead to less \edit{precise} (more statistical noise) results. From Fig. \ref{fig:sigmar}, it is evident that in a growing population regime (``Super''), the larger the ratio $N/M$, the larger the uncertainty introduced by the PCTs; this trend generally extends to the decaying population regime (``Sub''). However, some methods \edit{(SR, CO, and DD)} take advantage of the pure-splitting scenario---in which $M$ is a multiple of $N$---such that $\boldsymbol{\sigma_r}[C_i]$ drops to zero. \edit{We note that given a reasonably accurate prediction of population decay rate, one can take advantage of this behavior to minimize uncertainty introduced by the PCTs in systems with decaying population.} In terms of $\boldsymbol{\sigma_r}[C_i]$, SS is the worst PCT, followed by COX; particularly, unlike the other PCTs, SS and COX introduce significant uncertainties even when $N \approx M$ (which is the case throughout the active cycles of an eigenvalue simulation, see Sec. \ref{sec:test_eigen}). On the other hand, SR and CO are identically the best.

\subsection{Numerical Verification}\label{sec:test_verify}

To numerically verify the theoretical $\boldsymbol{\sigma_r}[C_i]$ derived in the previous subsection, we implement the PCTs into a Python-based research MC code and devise a PCT test problem. Per discussion in Sec. \ref{sec:samplingbasis}, only uniform PCTs are discussed here. Nevertheless, a similar set up can be used to verify PCTs with the other sampling bases.

In the test problem, we perform population control to an initial population with a cosine statistical weight distribution:
\begin{equation}
  w_i = \cos\left(\frac{i-1}{N-1}\pi\right)+1,\qquad i=1, 2, ..., N.
\end{equation}
Each initial particle $i$ is associated to tally bin $i$. All copies of particle $i$ in the final population will score their controlled weight $w_i'$ to the tally bin $i$; in other words, we are tallying $C_i$. 

\begin{figure}[H]
    \centering
    \subfloat[Simple Sampling]{%
    \resizebox*{7cm}{!}{\includegraphics{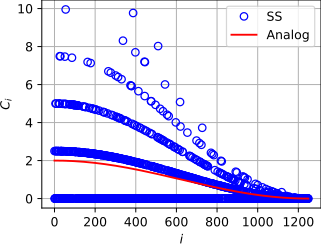}}}\hspace{5pt}
    \subfloat[Splitting-Roulette]{%
    \resizebox*{7cm}{!}{\includegraphics{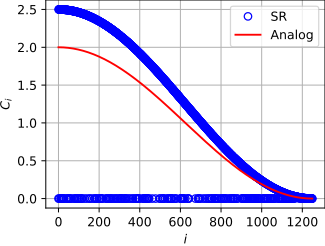}}}
    
    \subfloat[Combing]{%
    \resizebox*{7cm}{!}{\includegraphics{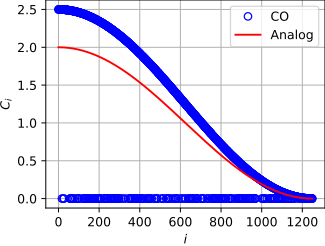}}}\hspace{5pt}
    \subfloat[Modified Combing]{%
    \resizebox*{7cm}{!}{\includegraphics{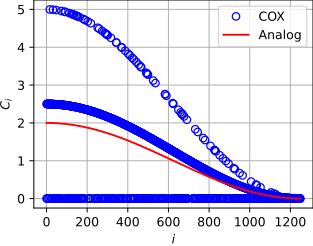}}}
    
    \subfloat[Duplicate-Discard]{%
    \resizebox*{7cm}{!}{\includegraphics{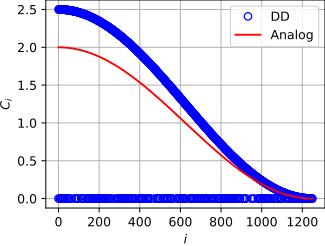}}}
    \caption{PCT test problem results with $N=1250$ and $M=1000$ of different techniques.} \label{fig:demo_super}
\end{figure}

\begin{figure}[H]
    \centering
    \subfloat[Simple Sampling]{%
    \resizebox*{7cm}{!}{\includegraphics{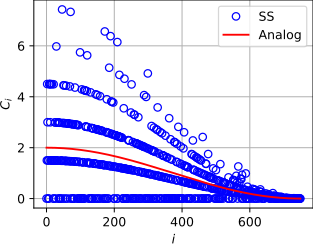}}}\hspace{5pt}
    \subfloat[Splitting-Roulette]{%
    \resizebox*{7cm}{!}{\includegraphics{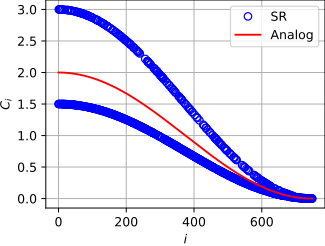}}}
    
    \subfloat[Combing]{%
    \resizebox*{7cm}{!}{\includegraphics{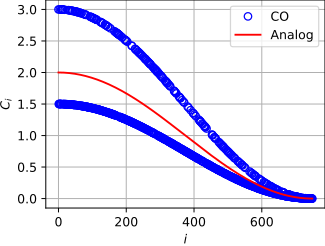}}}\hspace{5pt}
    \subfloat[Modified Combing]{%
    \resizebox*{7cm}{!}{\includegraphics{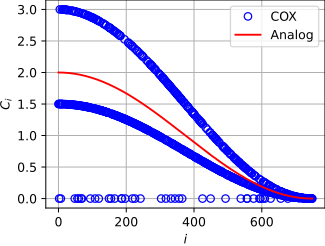}}}
    
    \subfloat[Duplicate-Discard]{%
    \resizebox*{7cm}{!}{\includegraphics{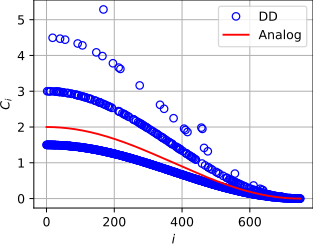}}}
    \caption{PCT test problem results with $N=750$ and $M=1000$ of different techniques.} \label{fig:demo_sub}
\end{figure}

Figs. \ref{fig:demo_super} and \ref{fig:demo_sub} show the resulting $C_i$ of different PCTs for $N/M=1.25$ and $N/M=0.75$, respectively. \edit{In each subplot, the red line indicates the analog result, where no PCT is performed and no uncertainty is introduced to the population ($C_i=w_i$). We note that there are $N$ discrete values of analog $C_i$, but we present it as a line to distinguish it with the values calculated using PCT, which are marked by the blue circles. As discussed earlier in this section, PCTs introduce some uncertainties to the population, and the magnitudes of the uncertainties are illustrated by how far the blue circles deviate from the red line: the more spread away the blue circles are from the red line, the more uncertainties are introduced by the respective techniques.} We note that the results shown in Figs. \ref{fig:demo_super} and \ref{fig:demo_sub} are in agreement to the theoretical uncertainty shown in Fig. \ref{fig:sigmar}---i.e, SS introduces the most uncertainty, followed by COX (and DD, for $N/M<1$), while CO and SR introduce the least.

\begin{figure}[H]
    \centering
    \includegraphics[width=0.5\textwidth]{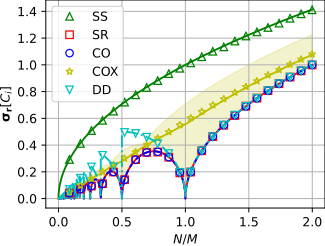}
    \caption{Verification of relative uncertainty $\boldsymbol{\sigma_r}[C_i]$ introduced by the PCTs.}
    \label{fig:sigmar_verify}
\end{figure}

\edit{Next, we would like to use the PCT test problem to verify the theoretical PCT uncertainties $\boldsymbol{\sigma_r}[C_i]$ derived in this section.} We set the target size $M$ to be 1000 and consider multiple values of $N$ such that $N/M$ ranges from 0.75 to 1.25. In each case, the population control is repeated 100 times so that we can determine the relative standard deviation $\boldsymbol{\sigma_r}[C_i]$ based on the accumulation of $C_i$ and $C_i^2$. Furthermore, we randomize the particle order in the population ``stack'' at each repetition. In uniform PCTs, $\boldsymbol{\sigma_r}[C_i]$ is independent of $i$ as it only depends on the value of $N/M$, as shown in Fig. \ref{fig:sigmar}. Therefore, in each case of $N/M$, we take the average of $\boldsymbol{\sigma_r}[C_i]$ over all $i$ as the final result. Finally, these numerical results from all cases of $N/M$ are compared to the theoretical values, as shown in Fig. \ref{fig:sigmar_verify}. The numerical results are denoted by the markers, and the lines are the theoretical values identical to those in Fig. \ref{fig:sigmar}. Excellent agreement is observed, even for COX with its ranging theoretical $\boldsymbol{\sigma_r}[C_i]$ (the shaded area). This verifies not only the theoretical $\boldsymbol{\sigma_r}[C_i]$ derived in Sec. \ref{sec:sigmar_theory}, but also the PCT implementations.

\section{Parallel PCT Algorithm}\label{sec:parallel}

Romano and Forget \cite{romano2012parallelBankAlg} introduce an efficient, \emph{reproducible}, parallel fission bank algorithm for $k$-eigenvalue MC simulation; in the paper, the typical uniform SS is described as the PCT. However, per our discussion in Secs. \ref{sec:census} and \ref{sec:popctrl}, we can actually apply the algorithm not only to the $k$-eigenvalue MC simulation (fission census) but also to the time-dependent fixed-source with time census. This allows us to adapt the algorithm and design a common population control code routine for both simulation modes. Furthermore, the PCT of choice can be any of the five PCTs discussed in Sec. \ref{sec:pcts}. \edit{In this section, we describe the adaptation of the parallel particle bank algorithm proposed in \cite{romano2012parallelBankAlg}, present the resulting pseudo-codes of the different PCTs, and perform a weak scaling study to their implementations.}

\begin{figure}[H]
    \centering
    \includegraphics[width=1\textwidth]{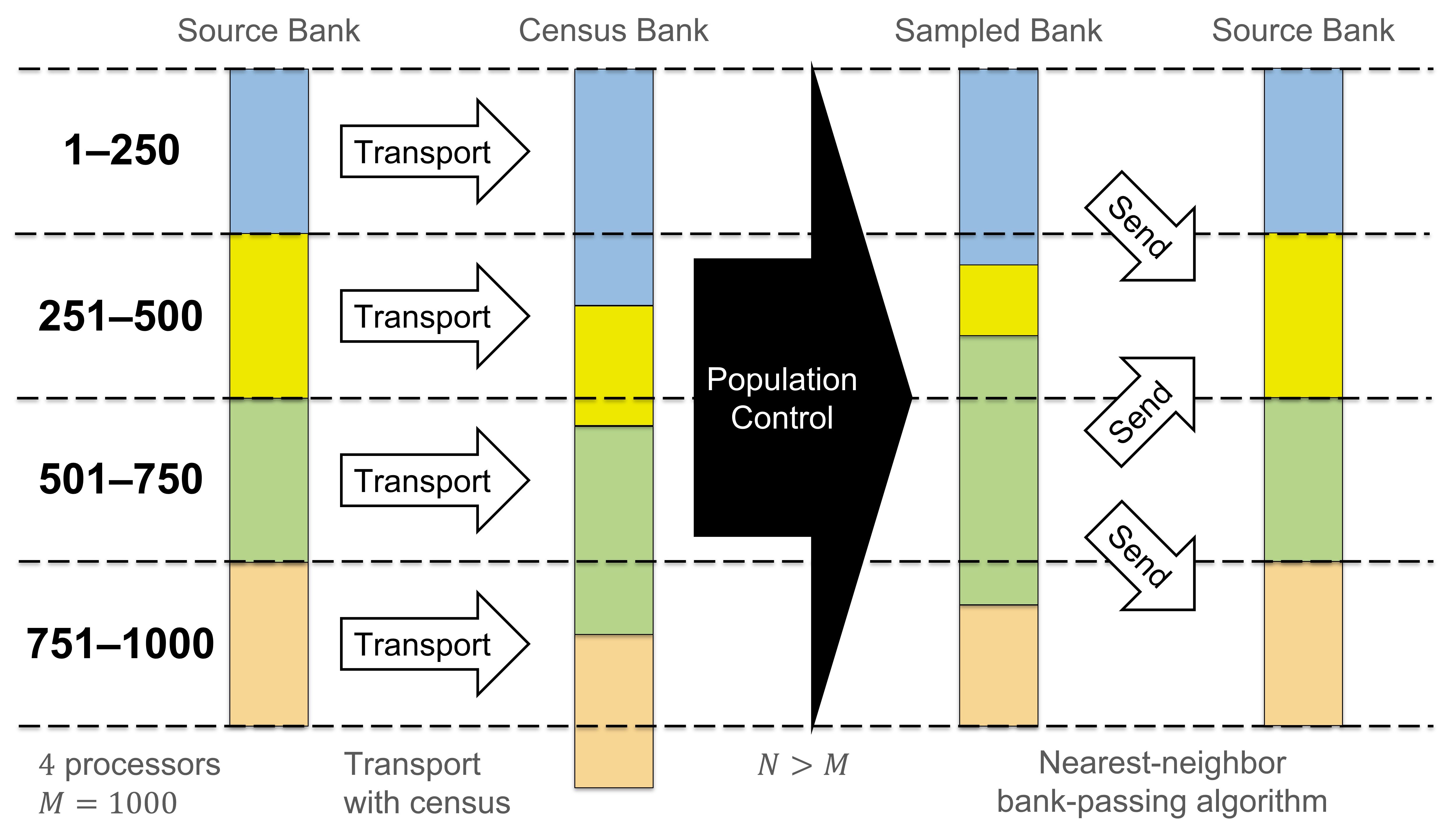}
    \caption{Illustration (adapted from \cite{romano2012parallelBankAlg}) of parallel particle bank handling and population control of the proposed algorithm.}
    \label{fig:parallel}
\end{figure}

Generalized from Fig. 3 in \cite{romano2012parallelBankAlg}, Fig. \ref{fig:parallel} illustrates an example of how particle banks are managed and population controlled using the proposed parallel algorithm. In the example, we consider 1000 source particles evenly distributed to 4 processors---each processor holds a Source Bank of size 250. The source particles are then transported in parallel. The transported particles are subject to a census mechanism, which can be a time census for time-dependent simulation or fission census for eigenvalue one. Once the particle census is completed, population control is performed to the Census Bank using one of the PCTs (SS, SR, CO, COX, or DD). Finally, the resulting final population (Sampled Bank) is evenly redistributed to the processors via the nearest-neighbor bank-passing algorithm, where each processor only needs to communicate (send or receive) with its adjacent neighbors as needed, without any global particle bank formation nor typical master-slave communication \cite{romano2012parallelBankAlg}.

Two exclusive scans need to be performed in the proposed parallel algorithm. An exclusive scan to the Census Bank is required to determine the total size $N$ and the position of the processor's local bank relative to the ``global'' bank, so that reproducible population control, regardless of the number of processors, can be achieved by consistently following the same random number sequence. The other scan is performed to the Sampled Bank so that we can determine local bank offsets required to perform the nearest-neighbor bank passing.

Algorithms \ref{alg:scan} and \ref{alg:pass} respectively show the pseudo-codes for bank-scanning and bank-passing processes, which are used in all of the PCT algorithms: Algs. \ref{alg:SS}--\ref{alg:DD}. The PCT algorithms only take the minimum information required to perform the population control---the Census Bank (which can be either fission or time bank) and the target size $M$---and return the controlled, evenly distributed across processors, final bank. Therefore, the proposed parallel PCT algorithms are applicable for both time-dependent fixed-source and eigenvalue MC simulation modes. We also note that the algorithms are designed to start and return with the same random number seed across all processors, which is important for maintaining reproducibility.

The parallel algorithms are implemented to the Python-based MC research code by using Python Abstract Base Class feature to allow streamlined implementation of the different PCTs---SS, SR, CO, COX, and DD. The distributed-memory parallel communication is facilitated by using MPI4Py \cite{mpi4py}. We use the verification test problem in Sec. \ref{sec:test_verify} to verify that the PCTs are properly implemented and their results (distribution of $C_i$) are reproducible---i.e., same results are produced regardless of the number of processors.

Next, we perform a weak scaling test to assess the relative parallel scalabilities of the different PCTs. The test is similar to the verification test problem in Sec. \ref{sec:test_verify}, except that $M$ is set to be $10^5$ times the number of processors, $N\in[0.5M,1.5M]$ is randomly picked in 50 repetitions, and the initial particles are randomly distributed to the processors. \edit{We note that there is no particle transport performed; we specifically measure the runtime required to perform population control and manage the particle bank, which includes communicating attributes and members of the bank throughout the processors.}

\begin{figure}[H]
    \centering
    \includegraphics[width=0.5\textwidth]{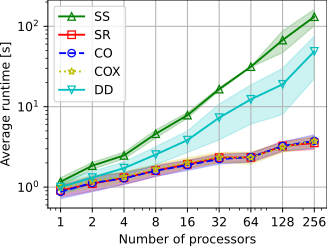}
    \caption{Weak scaling results of different PCTs. Marked solid lines and the associated shaded areas denote the average and standard deviation of the 50 repetitions, respectively.}
    \label{fig:scale}
\end{figure}

The weak scaling result is shown in Fig. \ref{fig:scale}. The marked lines and the shaded areas respectively show the average and the standard deviation of the runtimes in the 50 repetitions. It is found that SR, CO, and COX identically scale the best, followed by DD, and then SS. The sampling mechanisms of SR, CO, and COX are embarrassingly parallel, which make the techniques scale very well. On the other hand, SS, which scales the worst, needs to serially sample all of the $M$ particles. The sampling mechanism of DD is also done in serial, but it only needs to sample as many as the difference between $N$ and $M$; this makes DD scale better than SS, and also explains why DD's runtime has a relatively larger standard deviation. \edit{Finally, it is worth mentioning that we purposely picked a modest number of particles per processor ($10^5$) to get a balanced demonstration on the significance of work (population control sampling) and communication. Should we significantly increase the number of particles per processor, such that the amount of work far outweighs the amount of communication, parallel scalabilities of SR, CO, and COX will improve, approaching the perfect scaling (horizontal line).}

\section{Time-Dependent Problems}\label{sec:test_td}

In this section, we devise time-dependent MC test problems and then solve them with the PCTs to assess their relative performances. We adapt the homogeneous infinite 1D-slab medium problem of the analytical time-dependent benchmark suite AZURV1 \cite{ganapol2001azurv1}:
\begin{equation}\label{eq:azurv1}
  \left[ \frac{\partial}{\partial t} + 
         \mu \frac{\partial}{\partial x} + 
         1 \right] \psi(x,\mu,t) = \frac{c}{2} \phi(x,t) 
         + \frac{1}{2} \delta(x) \delta(t),
\end{equation}
which is subject to the following \edit{boundary and initial} conditions:
\begin{equation}
  \lim_{|x|\to\infty} \psi(x,\mu,t) < \infty, \quad \quad
  \psi(x,\mu,0)=0.
\end{equation}
Note that particle position and time are respectively measured in mean-free-path ($\Sigma_t^{-1}$) and mean-free-time [$(v\Sigma_t)^{-1}$] where $v$ is particle speed; and we also have the typical scattering parameter $c=(\Sigma_s +\nu\Sigma_f)/\Sigma_t$. The scalar flux solution \(\phi(x,t)=\int_{-1}^{1} \psi(x,\mu,t) \,d\mu\) of this time-dependent problem is
\begin{equation}\label{eq:azurv_phi}
  \phi(x,t) = 
    \frac{e^{-t}}{2t} \left\{1 + \frac{ct}{4\pi} \left(1-\eta^2\right) 
    \int_{0}^{\pi} \sec^2\left(\frac{u}{2}\right)
    \boldsymbol{Re}\left[\xi^2e^{\frac{ct}{2}\left(1-\eta^2\right)\xi}\right] \,du \right\} 
    \boldsymbol{H}(1-|\eta|),
\end{equation}
where 
\begin{equation}
  \eta = \frac{x}{t}, \quad \quad q = \frac{1+\eta}{1-\eta},
  \quad \quad \xi(u) = 
    \frac{\ln(q) + iu}{\eta + i \tan\left(\dfrac{u}{2}\right)},
\end{equation}
and $\boldsymbol{H}(\cdot)$ denotes the heaviside function.

\begin{figure}[H]
    \centering
    \subfloat[Supercritical, $c=1.1$]{%
    \resizebox*{7cm}{!}{\includegraphics{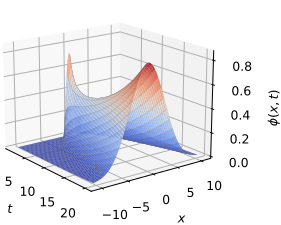}}}\hspace{5pt}
    \subfloat[Subcritical, $c=0.9$]{%
    \resizebox*{7cm}{!}{\includegraphics{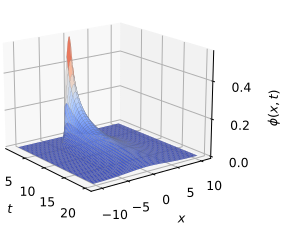}}}
    \caption{Reference solution of the time-dependent test problems.} \label{fig:td_analytical}
\end{figure}

For our test problems we consider $c$ values of 1.1 and 0.9, respectively representing supercritical and subcritical systems. The analytical solution of the total flux would be a simple exponential function of $\phi(t) = \exp{\left[(c-1)t\right]}$; however, the spatial solutions [Eq. \eqref{eq:azurv_phi}] offer some more interesting features, particularly for the supercritical case, as shown in Fig. \ref{fig:td_analytical} (note that the solutions in $t\leq1$ and $|x|\in[10,20]$ are not shown to better show the prominent spatial features).

The test problems are initiated by an isotropic neutron pulse at $x=t=0$ (c.f. \eqref{eq:azurv1}). In both cases, the scalar flux solution gradually diffuses throughout the medium. The difference is that the significant neutron absorption promotes population decay in the subcritical case. On the other hand, while the solution of the supercritical case initially behaves similarly to that of the subcritical, it eventually raises up due to the significant fission multiplication---at $t=20$, the population size reaches $\exp(2)=7.39$ times of the initial value.

\subsection{Verifying Time-Dependent Features of the MC code}\label{sec:td_verif}

\edit{To the authors' knowledge, there are three different time-dependent scalar flux quantities that can be calculated via MC simulation: (1) spatial-average time-average $\phi_{j,k}$, (2) spatial-average time-edge $\phi_j(t)$, and (3) spatial-edge time-average $\phi_k(x)$, where $j$ and $k$ respectively denote spatial and time mesh indices.} The first tally uses the typical track-length estimator averaged over time mesh. The second uses a time-edge estimator, which accumulates the product of neutron speed and weight whenever a time-grid is crossed, averaged over spatial mesh. The third uses the typical spatial-mesh-crossing estimator, which scores particle weight divided by absolute of normal product of particle direction and the surface, averaged over time mesh. \edit{The use of the track length estimator (spatial-average time-average $\phi_{j,k}$) is typically desired because it generally gets more samples compared to ``event-triggered" estimators like time-edge-crossing ($\phi_j(t)$) and spatial-mesh-crossing ($\phi_k(x)$). Nevertheless, it is worth mentioning that in some applications, time-edge quantities $\phi_j(t)$ may be more desired than the time-average one $\phi_{j,k}$.}

\begin{figure}[H]
    \centering
    \includegraphics[width=0.5\textwidth]{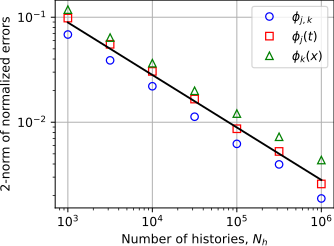}
    \caption{Error convergence of the three time-dependent flux tallies of analog (without PCT) MC simulations of the supercritical test problem. Black solid line indicates convergence rate of $O(1/\sqrt{N_h}$).}
    \label{fig:td_analog}
\end{figure}

To simulate the supercritical ($c=1.1$) and subcritical ($c=0.9$) cases, we consider purely fission media with $\nu=c$. The test problems are simulated using the research MC code, and we record the scalar flux using the three tally estimators that are subject to $J=202$ uniform spatial meshes spanning $x\in[-20.5,20.5]$ and time grid $t=0, 1, 2, ..., 20$. To limit particle population growth in the supercritical case, we set a time boundary at the final time $t=20$---particles crossing this time boundary will be killed (analogous to spatially crossing a convex vacuum boundary). Note that we have not introduced any PCT yet; the MC simulation is still run in analog. Simulations are performed with increasing number of histories $N_h$. The resulting 2-norms of normalized error [against the reference formula Eq. \eqref{eq:azurv_phi}, normalized at each time index] of the supercritical problem are shown in Fig. \ref{fig:td_analog}. It is found that all of the error 2-norms converge at the expected rate of $O(1/\sqrt{N_h})$ (shown in black solid line); and the track length estimator result $\phi_{j,k}$ shows the lowest error, which is in line with the discussion in the previous paragraph. Similar convergence rate is observed in the subcritical case as well. This verifies the time-dependent features of the MC code that we are going to use in the next subsection to assess the relative performances of the PCTs. Additionally, this also suggests that this set of test problems, the AZURV1 benchmark \cite{ganapol2001azurv1}, serve as a good verification tool to test time-dependent features of MC codes.

\subsection{Performances of the PCTs in solving the Time-Dependent Test Problems}\label{sec:td_pct}

The supercritical and subcritical problems are solved using the 5 PCTs (SS, SR, CO, COX, DD). Each simulation is run with $10^5$ source particles on 36 distributed-memory processors. We consider uniformly-spaced population control time censuses within $t\in[0,20]$. With increasing frequency, we consider 8 number of censuses: 1, 2, 4, 8, 16, 32, 64, 128. In 1 census, the census is performed at $t=10$; while in 2 censuses, it is performed at $t=20/3$ and $t=40/3$. Finally, each simulation is repeated 100 times with different random number seeds.

\begin{table}[H]
\singlespacing
\tbl{Census configurations for the time-dependent test problems.}
{\begin{tabular}{llcccccccc} 
 \toprule
  \multicolumn{2}{l}{Number of censuses in $t\in[0,20]$} & 1 & 2 & 4 & 8 & 16 & 32 & 64 & 128 \\
\midrule
  \multicolumn{2}{l}{Census period (mean-free-time)} & 10.0 & 6.67 & 4.00 & 2.22 & 1.18 & 0.61 & 0.31 & 0.16 \\
\midrule
  \multirow{2}{*}{Expected $N/M$} & Supercritical & 2.72 & 1.95 & 1.49 & 1.25 & 1.12 & 1.06 & 1.03 & 1.02 \\
  & Subcritical & 0.37 & 0.51 & 0.67 & 0.8 & 0.89 & 0.94 & 0.97 & 0.98 \\
 \bottomrule
\end{tabular}}
\label{tab:census}
\end{table}

\begin{figure}[H]
    \centering
    \resizebox*{6.5cm}{!}{\includegraphics{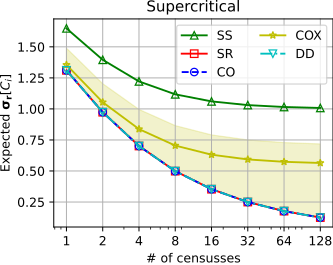}} \hspace{5pt}
    \resizebox*{6.5cm}{!}{\includegraphics{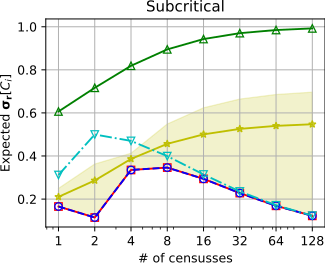}}
    \caption{\edit{Expected $\boldsymbol{\sigma_r}[C_i]$ of the simulated cases (c.f. Table \ref{tab:census} and Fig. \ref{fig:sigmar}).}} \label{fig:td_sigmar}
\end{figure}

Table \ref{tab:census} shows census period and the expected ratio $N/M$ associated with the simulated cases. By referring to Fig. \ref{fig:sigmar}, we can estimate the expected uncertainty $\boldsymbol{\sigma_r}[C_i]$ introduced by a PCT at a given value of $N/M$. \edit{For convenience, plots showing the expected $\boldsymbol{\sigma_r}[C_i]$ associated to the simulated cases are provided in Fig. \ref{fig:td_sigmar}}. Note that this expected uncertainty is introduced every time the population control is performed---e.g., with 4 number of censuses, we perform census and population control, and introduce the associated uncertainty, once every 4 mean-free-times. This means, smaller $\boldsymbol{\sigma_r}[C_i]$ due to larger census frequency does not necessarily lead to smaller uncertainty in the simulation result, because the more frequently we perform population control, the more frequently we introduce the uncertainty $\boldsymbol{\sigma_r}[C_i]$ (even though small) to the population.

\edit{Two performance metrics are considered: (1) total runtime $T$ and (2) the averaged sample standard deviation of the scalar flux at the end of simulation time $\phi_j(t=20)$, or simply $\phi_j$. The sample standard deviation $\boldsymbol{\sigma}[\phi_j]$ (not the mean standard deviation) is calculated over the $N_r=100$ repetitions:}
\begin{equation}
    \bar{\phi}_j = \frac{1}{N_r}
        \sum_{i=1}^{N_r}\phi_j^{(i)},
\end{equation}
\begin{equation}
    \boldsymbol{\sigma}[\phi_j]^2 = \frac{1}{N_r-1}
        \left[\sum_{i=1}^{N_r}\left[\phi_j^{(i)}\right]^2-N_r\bar{\phi}_j^2\right],
\end{equation}
\edit{where the superscript $^{(i)}$ denotes repetition index. We note that a repetition (or a realization) can be seen as a batch of source particles that makes a single, independent history. Finally, the averaged sample standard deviation of the scalar flux is calculated as follows:}
\begin{equation}\label{eq:sdv}
    \boldsymbol{\bar{\sigma}}[\phi] = \frac{1}{J}
        \sum_{j=1}^{J}\boldsymbol{\sigma}[\phi_j].
\end{equation}

\begin{figure}[H]
    \centering
    \subfloat[Runtime]{%
    \resizebox*{6.5cm}{!}{\includegraphics{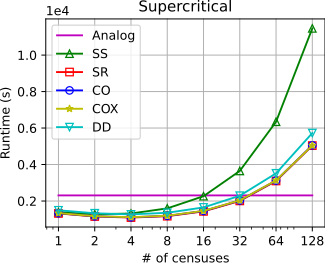}}
    \hspace{5pt}
    \resizebox*{6.5cm}{!}{\includegraphics{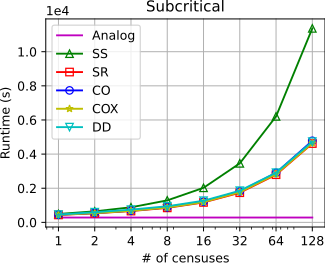}}}

    \subfloat[\edit{Averaged sample standard deviation of the scalar flux}]{%
    \resizebox*{6.5cm}{!}{\includegraphics{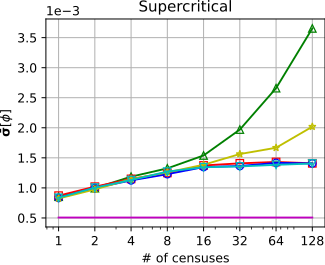}}
    \hspace{5pt}
    \resizebox*{6.5cm}{!}{\includegraphics{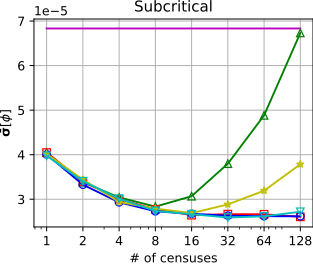}}}
    
    \subfloat[Figure of merit]{%
    \resizebox*{6.5cm}{!}{\includegraphics{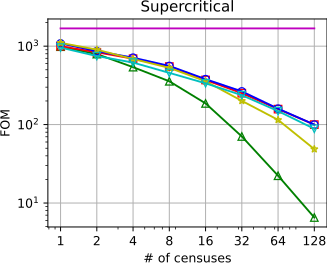}}
    \hspace{5pt}
    \resizebox*{6.5cm}{!}{\includegraphics{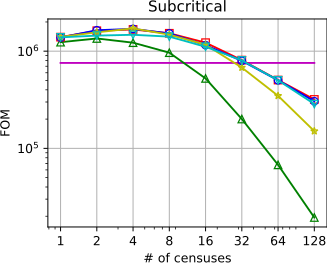}}}
    \caption{Performance metrics of different PCTs for the time-dependent problems.} \label{fig:td_pct}
\end{figure}

\edit{The resulting performance metrics are shown in Fig. \ref{fig:td_pct}. A figure of merit (FOM) based on the two performance metrics,}
\begin{equation}\label{eq:fom}
  FOM = \frac{1}{T\boldsymbol{\bar{\sigma}}[\phi]^2} ,
\end{equation}
\edit{is also shown in the figure. Finally, the analog (without PCT) solution, also run in 100 repetitions, is shown in the figure as well as a reference point. Figure \ref{fig:td_pct} not only compares the relative performance of the PCTs but also shows the trends of the related metrics as functions of census frequency. The figure also illustrates how PCT functions differently in supercritical and subcritical problems.}

\subsubsection{Supercritical problem}

The main motivation of population control in a supercritical problem is to limit the number of neutrons tracked during the simulation so that it does not exceed the allocated computational memory---in the test problem, population size exceeds seven times of the initial value if population control is not performed. However, this comes at the expense of less \edit{precise} (more noisy) solution due to the significant uncertainty introduced by the PCT used.

Figure (a) of Fig. \ref{fig:td_pct} shows that applying PCT in a supercritical problem potentially reduces the overall runtime. However, too frequent census may result to net increase in runtime (relative to analog) due to the significant cost of performing too many population controls, which may involve considerable parallel communications. The figure also shows that SR, CO, and COX have the lowest runtime, followed by DD and then SS, which is in agreement with the discussion in Sec. \ref{sec:parallel}.

Figure (b) of Fig. \ref{fig:td_pct} shows the averaged scalar flux standard deviations $\boldsymbol{\bar{\sigma}}[\phi]$ of the different PCTs as functions of number of censuses performed. The averaged scalar flux standard deviation is a measure of how noisy the simulation is---the larger $\boldsymbol{\bar{\sigma}}[\phi]$, the lower the simulation precision and the larger the noise in the result. The figure demonstrates the significance of the uncertainty introduced by the PCTs (note the lower $\boldsymbol{\bar{\sigma}}[\phi]$ value of the analog result). Generally, the more frequently we perform population control, the more uncertainty is introduced to the population, and the larger $\boldsymbol{\bar{\sigma}}[\phi]$. While $N/M$ (as well as $\boldsymbol{\sigma_r}[C_i]$, per Figs. \ref{fig:sigmar} and \ref{fig:td_sigmar}) reduces as we increase the census frequency, the number of population controls performed and thus how often the uncertainty is introduced also increase. It is shown that all PCTs seem to yield similar averaged scalar flux standard deviations in the lower census frequency. However, as we increase the census frequency, SR, CO, and DD seem to limit their standard deviations; this demonstrates their superiority over COX and SS as the three techniques theoretically introduce the least uncertainty in supercritical problems, as shown in Figs. \ref{fig:sigmar} and \ref{fig:td_sigmar}.

Finally, figure (c) of Fig. \ref{fig:td_pct} shows that the FOMs of all PCTs are always lower than that of the analog simulation, and they monotonically decrease as we increase the census frequency\edit{---it seems that PCT is parasitic in this MC simulation.} However, we should note that the main reason of applying PCT in a supercritical system is to limit population size being tracked in the simulation. \edit{Nevertheless, in general, one may find situations where the PCTs have larger FOMs than the analog one (for smaller census frequency) should the advantage of runtime reduction significantly outweighs the uncertainty introduced by the PCTs.} Another important takeaway from the figure is that SR, CO, and DD are in the same ballpark as the best PCTs, which are followed by COX, and then SS.

\subsubsection{Subcritical problem}

The main motivation of population control in a subcritical problem is to maintain population size so that we have enough samples to yield more precise (less noisy) solution. However, this comes at the expense of increasing overall runtime as more neutrons need to be tracked.

Figure (a) of Fig. \ref{fig:td_pct} shows that applying PCT in a subcritical problem increases overall runtime, and it increases further as we perform the population control more frequently. It is also worth to mention that DD has a similar runtime to those of SR, CO, and COX in higher number of censuses; this is because DD only needs to sample as many as $|N-M|$, which gets closer to zero as we increase census frequency. 

Figure (b) of Fig. \ref{fig:td_pct} shows that population control improves the solution precision. One may think that the solution would improve further as the population control is performed more frequently; however, we should be aware that population control introduces uncertainty in a subcritical problem too (see Figs. \ref{fig:sigmar} and \ref{fig:td_sigmar}). The effect of this uncertainty is evident in the figure (b) of Fig. \ref{fig:td_pct}---at around 8 censuses, the solution improvement starts to deplete, and even reversed ($\boldsymbol{\bar{\sigma}}[\phi]$ increases) for SS and COX.

Finally, figure (c) of Fig. \ref{fig:td_pct} shows that the PCTs offer improved FOMs relative to the analog. The FOM is improved further as we perform population control more frequently. However, it starts to consistently degrade as the effects of the increasing runtime and of the significant uncertainty introduced by the PCT start to dominate. Note that this is similar to the typical trend of a variance reduction technique: it helps to improve FOM, but will degrade FOM if it is used too much. Another important takeaway from the figure (c) of Fig. \ref{fig:td_pct} is that---similar to the supercritical case---SR, CO, and DD are in the same ballpark as the best PCT, followed by COX, and then SS.

\section{$k$-Eigenvalue Problem}\label{sec:test_eigen}

\edit{In a MC calculation, the $k$-eigenvalue transport problem is typically solved via the method of successive generations. The MC simulation involves accumulation of fission neutrons in a fission bank. At the end of each generation (i.e., when the fission census is completed), the eigenvalue $k$ is updated, and the generated fission bank is normalized such that its total weight is identical to the target population size $M$, which is the number of histories per generation. Finally, the normalized fission bank is set to be the source bank for the subsequent generation.}

\edit{The procedure described above is considered to be the ``analog" approach where there is no PCT being used. The eigenvalue update and weight normalization are useful for the MC simulation, because they direct the neutronics system into the steady-state configuration which helps maintaining the number of source particles simulated at each generation around the user-specified value $M$. However, in the earlier generations, when the steady-state configuration---i.e. ``convergence" of the eigenvalue $k$---has not been achieved yet, this analog technique may suffer from highly-fluctuating number of source
particles, particularly if the initial guesses for the eigenvalue $k$ and the “eigenvector” neutron distribution are poorly chosen. This possible issue can be avoided by performing population control (applying one of the identified PCTs) to the normalized fission bank, so that the resulting source bank is well controlled, regardless of the convergence of the eigenvalue $k$.}

It is worth mentioning that the ``eigenfunction normalization" described in the previous paragraphs and the ``PCT normalization" discussed in Sec. \ref{sec:normalization} serve different purposes. The eigenfunction normalization is a necessary step to ensure that scores accumulated into simulation tallies are not arbitrary in magnitude. On the other hand, PCT normalization is an optional step to preserve the total weight of the initial population passed to the PCT (at the expense of introducing bias, as discussed in Sec. \ref{sec:normalization}). As another clear distinction, the eigenfunction normalization is performed before we apply PCT, while the optional PCT normalization is performed after.

\edit{Similar to the PCT application in time-dependent MC simulation, PCT application in $k$-eigenvalue simulation would also introduce the uncertainty $\boldsymbol{\sigma_r}[C_i]$ to the population. However, it should be emphasized that this uncertainty introduced by PCT is not a bias. There is a well-known bias associated with the method of successive generations (which can be mitigated given sufficiently large number of histories per generation \cite{brown2009}). This bias is introduced when the fission bank is normalized, which is in effect regardless whether population control is applied. PCT neither enhances nor eliminates this bias---it merely introduces the additional uncertainty $\boldsymbol{\sigma_r}[C_i]$ to the already-bias simulation.}

\edit{In an eigenvalue simulation, how many times population control is performed is determined by the total number of generations, which is typically a very large number (in the order of $10^2$ to $10^3$). This means, the uncertainty $\boldsymbol{\sigma_r}[C_i]$ would be introduced by the PCT to the population so many times, which, according to the findings in the previous section, may lead to highly noisy solutions. However, the effect of the uncertainty introduced by the PCT on an eigenvalue simulation is expected to be much less pronounced than that in a time-dependent one. This is because once the eigenvalue convergence is achieved, we essentially simulate a steady-state system, where the ratio $N/M$ is expected to be around and close to unity, in which most PCTs introduce minimum uncertainties, as shown in Fig. \ref{fig:sigmar}. Nevertheless, some PCTs (SS and COX) still introduce considerably high uncertainties even with $N/M\approx1$. In this section we would like to investigate the relative performances of the identified PCTs in a $k$-eigenvalue transport calculation---particularly, we would like to see whether there is any discernible effect due to the different magnitudes of uncertainty $\boldsymbol{\sigma_r}[C_i]$ introduced by the techniques.}

\begin{figure}[H]
    \centering
    \includegraphics[width=0.5\textwidth]{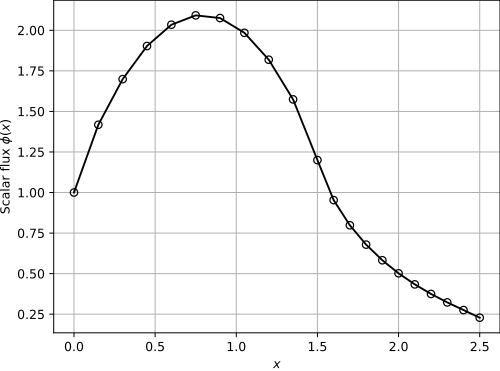}
    \caption{Scalar flux associated with the fundamental $k$-eigenvalue of the test problem \cite{kornreich2004k}.}
    \label{fig:eigenvalue_problems}
\end{figure}

We consider the $k$-eigenvalue transport problem of the mono-energetic two-region slab medium from \cite{kornreich2004k}:
\begin{equation}
   \left[ \mu\frac{\partial}{\partial x} + \Sigma_t(x) \right]\psi(x,\mu) = \frac{1}{2}\left[\Sigma_s(x) + \frac{1}{k}\nu\Sigma_f(x) \right]\phi(x),
\end{equation}
Similar to Sec. \ref{sec:test_td}, all physical quantities will be presented in the unit of mean-free-path. The first and the second regions respectively occupy $x\in[0,1.5]$ and $x\in[1.5,2.5]$. The cross-sections of the two regions are $\nu\Sigma_{f,1}=0.6$, $\Sigma_{s,1}=0.9$, $\nu\Sigma_{f,2}=0.3$, and $\Sigma_{s,2}=0.2$. Finally, the two-region slab is subject to vacuum boundaries. By using a deterministic transport method, Kornreich and Parsons \cite{kornreich2004k} provide reference values for the fundamental $k$-eigenvalue, $k=1.28657$, and the associated scalar fluxes at certain points (shown in Fig. \ref{fig:eigenvalue_problems}).

\edit{The $k$-eigenvalue problem is solved using the analog (without population control) weight normalization technique and the five identified PCTs (SS, SR, CO, COX, and DD). The numbers of inactive and active generations are set to be 100 and 200, respectively, with $10^5$ neutron histories per generation. We tally the spatial-average neutron flux $\phi_j$ with $J=50$ uniform meshes spanning $x\in[0,2.5]$. Uniform isotropic flux distribution and $k=1$ are used as the initial guess. Finally, each simulation is repeated 50 times with different random number seeds and run on 36 distributed-memory processors. Solution of each run is verified by comparing it with the reference solution shown in Fig. \ref{fig:eigenvalue_problems}.}

\begin{figure}[H]
    \centering
    \subfloat[(Left) averaged sample standard deviation of the scalar flux and (Right) sample standard deviation of the eigenvalue. The values are relative to the median of the analog technique.]{%
    \resizebox*{6.5cm}{!}{\includegraphics{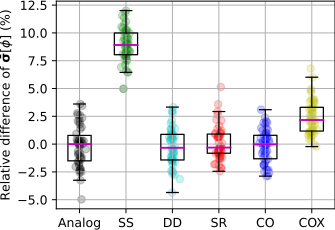}}
    \hspace{5pt}
    \resizebox*{6.5cm}{!}{\includegraphics{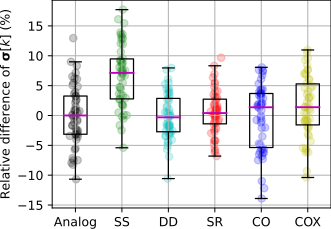}}
    }
    
    \subfloat[(Left) Runtime spent in population control and managing particle fission bank and (Right) total runtime.]{%
    \resizebox*{6.5cm}{!}{\includegraphics{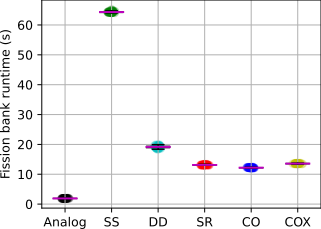}}
    \hspace{5pt}
    \resizebox*{6.5cm}{!}{\includegraphics{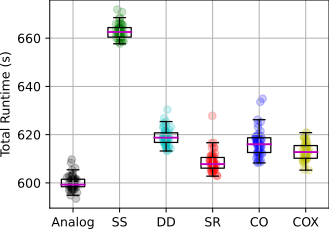}}}

    \subfloat[Figure of merit. The values are relative to the median of the analog technique.]{%
    \resizebox*{6.5cm}{!}{\includegraphics{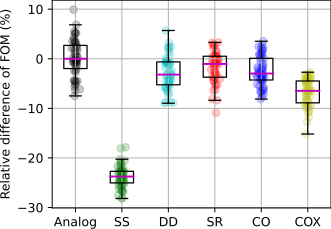}}}
    \caption{Performance metrics of different PCTs for the $k$-eigenvalue test problem. Magenta line indicates median; box indicates lower ($Q_1$) and upper ($Q_3$) quartiles; lower and upper whiskers indicate minimum and maximum values that are not outliers (within the range $\left[Q_1-1.5IQR, Q_3+1.5IQR\right]$, where $IQR=Q_3-Q_1$).} \label{fig:eigenvalue_pct}
\end{figure}

Three performance metrics are considered: (1) averaged sample standard deviation of the scalar flux $\boldsymbol{\bar{\sigma}}[\phi]$ (c.f. Eq. \eqref{eq:sdv}), (2) sample standard deviation of the eigenvalue $\boldsymbol{\sigma}[k]$, and (3) total runtime. The resulting performance metrics of the different PCTs are compared in jittered box plots shown in Fig. \ref{fig:eigenvalue_pct}.

\edit{The analog weight normalization technique is expected to have the least noisy solution. This is because, unlike the PCTs, the analog technique does not introduce any additional uncertainty to the population. The comparison presented in Fig. \ref{fig:eigenvalue_pct} helps in identifying the cost of performing population control (via one of the PCTs) instead of the analog weight normalization technique.

Part (a) of Fig. \ref{fig:eigenvalue_pct} compares the scalar flux averaged sample standard deviations $\boldsymbol{\bar{\sigma}}[\phi]$ and the eigenvalue sample standard deviations $\boldsymbol{\sigma}[k]$. The values are relative to the medians of the analog technique. Comparing $\boldsymbol{\bar{\sigma}}[\phi]$, figure on the left implies that SS yields considerably larger noise to the simulation, which makes $\boldsymbol{\bar{\sigma}}[\phi]$ about 8.5\% larger than the analog case. A discernible increase in $\boldsymbol{\bar{\sigma}}[\phi]$ is also found in COX, but only about 2\%. The other PCTs (DD, SR, and CO), however, does not really suffer from an increase in $\boldsymbol{\bar{\sigma}}[\phi]$. These findings are in agreement with the theoretical uncertainty introduced by the PCTs shown in Fig. \ref{fig:sigmar}. Similar trend can be observed in the figure on the right that compares $\boldsymbol{\sigma}[k]$, but it is not as pronounced since the data is widely fluctuating. This high fluctuation in $\boldsymbol{\sigma}[k]$ can be reduced by increasing the number of active generations or the number of particle histories per generation. We note that given the current configuration (200 generations, $10^5$ particles/generation), the median of $\boldsymbol{\sigma}[k]$ of the analog technique is 400.3 pcm, where the associated standard deviation of the expected mean is 28.3 pcm.

From the left figure of part (b) of Fig. \ref{fig:eigenvalue_pct}, it is shown that SS runs much slower than the other PCTs as it suffers from its serial particle sampling. DD also suffers from a serial sampling; but different to SS, the serial sampling of DD only needs to be done $|N-M|$ times, which is close to zero throughout the active generations of the simulation. This makes DD significantly faster than SS. On the other hand, SR, CO, and COX benefit from their embarrassingly parallel sampling procedures, which make them the fastest among the PCTs. Although not performing any population control, the analog case still spends an amount of time as it still needs to perform the bank passing procedure (Alg. \ref{alg:pass}). Finally, the figure on the right demonstrates that the fission bank handling and population control take a considerable portion of the total simulation runtime (about 9\% for SS, and 3\% for the other PCTs), which is enough to make SS about 10\% slower than the analog case; while the other PCTs are just about 1.5--3\% slower.

Finally, part (c) of Fig. \ref{fig:eigenvalue_pct} compares the resulting FOMs of the PCTs. The FOM follows the definition in Eq. \eqref{eq:fom}. It is found that SS---the simplest, yet the most well-known technique for $k$-eigenvalue simulation---is the least performant PCT having FOM about 24\% lower than the analog technique. On the other hand, SR, CO, and DD are the best PCTs with FOM about 2 to 4\% lower. However, it is worth mentioning that the discernible decrease in FOMs of SR, CO, and DD is due to their higher runtimes (right figure of figure (b)), since their sample standard deviations (figure (a)) are about the same as the one with the analog technique. The higher runtimes of the PCTs can be hidden should we run the simulations with a sufficiently large number of particles per processor. In such a situation, SR, CO, and DD would be as performant as the analog technique.}

\section{Conclusion}\label{sec:conclusion}

\edit{An extensive} study on population control technique (PCT) for time-dependent and eigenvalue Monte Carlo (MC) neutron transport calculations is presented. We define PCT as a technique that takes a censused population and returns controlled, unbiased one. A new perspective based on an abstraction of particle census and population control is explored, paving the way to improved understanding and application of the concepts. We discuss how different kinds of census---e.g., time, fission, and collision censuses---are performed in time-dependent and eigenvalue problems. We also discuss the requirements and desirable characteristics of a PCT.

Identified from the literature, five distinct PCTs are reviewed: Simple Sampling (SS), Splitting-Roulette (SR) \cite{sweezy2014mcatkSR}, Combing (CO) \cite{booth1996comb}, modified Combing (COX) \cite{ajami2021newCombing}, and Duplicate-Discard (DD) \cite{leppanen2013serpent2TDMC}. While SS has been the typical procedure of choice in handling fission bank in MC eigenvalue simulations, the other four techniques have been almost exclusively applied for time-dependent simulation. The review encompasses the basic procedures of the techniques, significance of their sampling bases (uniform, weight-based, and importance-based), bias in PCT weight normalization, possible correlation issue in CO, thorough characterization of the recently introduced COX, and relation to the more advanced PCTs \cite{booth1996comb,legrady2020pctVR}. A short remark of the five PCTs, highlighting their respective caveats, are summarized in Table \ref{tab:summary}.

\begin{table}[h]
\tbl{Short remark on the five PCTs.}
{\begin{tabular}{ll} 
\toprule
 PCT & Remark \\
 \midrule
 SS & Low parallel scalability, introduce largest uncertainty to the population\\
 DD & Limited parallel scalability\\
 SR & \edit{Does not exactly (1) yield targeted population size and (2) preserve total weight}\\
 CO & Subject to possible undesirable behavior due to correlation in particle order \\
 COX & Avoid CO’s issue, at the expense of increased uncertainty introduced \\
 \bottomrule
\end{tabular}}
\label{tab:summary}
\end{table}

A theoretical analysis on the uncertainty introduced to population by each of the PCTs is presented. The resulting theoretical uncertainties (shown in Fig. \ref{fig:sigmar}) are useful not only for theoretically assessing the relative performance of the PCTs, but also for numerically verifying whether the techniques are appropriately implemented. It is found that CO and SR are equally the most performant techniques based on this metric (smallest uncertainty introduced), followed by DD, then COX, while SS introduces the largest uncertainty. We hypothesized that this uncertainty would proportionally affect simulation tally results, which was later confirmed when we ran test problems using the different PCTs.

Parallel algorithms for the five PCTs are proposed. The algorithms are based on a generalized version of the parallel fission bank algorithm \cite{romano2012parallelBankAlg} and designed to be applicable for both eigenvalue and time-dependent simulations. The use of abstract base class for streamlined implementations of the five PCTs is also suggested. Weak scaling results of the PCTs are performed to demonstrate the parallel scalability of the techniques. It is found that SS and DD have limited scalabilities due to their respective significant serial sampling procedures.

Supercritical and subcritical time-dependent test problems based on the analytical benchmark AZURV1 \cite{ganapol2001azurv1} are devised; we found that these test problems serve as a good benchmark suite for verifying time-dependent features of a MC code. With the test problems we not only compare the relative performances of the PCTs, but also demonstrate typical behaviors of the PCTs in supercritical and subcritical problems as a function of census frequency. Two performance metrics are considered, total runtime and averaged sample standard deviation of the scalar flux, which are aggregated into figure of merit (FOM). Similar analysis is performed to a $k$-eigenvalue test problem. We found that SR and CO are equally the most performant PCTs, closely followed by DD, and then COX.

The results of the time-dependent and $k$-eigenvalue test problems demonstrate the superiority of SR and CO. However, that does not mean that everyone would be confident in using one of those techniques in all of their MC simulations. This is particularly true since there is caveat for each of the PCTs, as summarized in \ref{tab:summary}. The proposed generalized and streamlined PCT parallel algorithm offers MC code developers a minimally-invasive way to implement all of the PCTs into their code and allow the code users to pick the technique themselves. Finally, it is worth mentioning that the proposed PCT parallel algorithm can be generally applied to other MC eigenvalue simulations, such as $c$- \cite{kiedrowski2012ceig} and $\alpha$-eigenvalues \cite{variansyah2020alpha}.

Future work includes implementing the generalized parallel PCT algorithm into a production MC code and assessing the relative performances of the PCTs in simulating more practical, multi-dimensional, continuous-energy problems. \edit{Additionally, investigating the interplay between PCT and variance reduction technique (VRT), which is briefly discussed in Sec. \ref{sec:samplingbasis}, can be a potential future work. Finally, a recent analysis of neutron clustering in $k$-eigenvalue transport calculation demonstrates that sampling source sites without replacement (which is a variation of DD) versus with replacement (which is SS) can greatly reduce the degree of clustering \cite{sutton2021}. Studying how the different PCTs (particularly those who are not included in \cite{sutton2021}, i.e., SR, CO, and COX) affect the clustering phenomenon can be a potential future work as well.}

\section*{Acknowledgements}

This work was supported by the Center for Exascale Monte-Carlo Neutron Transport (CEMeNT) a PSAAP-III project funded by the Department of Energy, grant number DE-NA003967.

\bibliographystyle{unsrt}
\bibliography{references}

\appendix
\section{Parallel Algorithms for the Population Control Techniques}

\begin{algorithm}[H] 
\singlespacing
\caption{Scanning ``global'' bank to get local position and bank sizes}
\label{alg:scan}
\begin{algorithmic}[1]
\Function{BankScanning}{bank}
  \State {N\_local = size(bank)}
  \State {idx\_start = MPI\_ExclusiveScan(N\_local)}
  \State {N\_global = \Call{MPI\_Broadcast}{idx\_start+N\_local, from=last\_rank}}
  \State \Return {idx\_start, N\_local, N\_global}
\EndFunction
\end{algorithmic}
\end{algorithm}

\begin{algorithm}[H]
\singlespacing
\caption{Nearest-neighbor bank passing}
\label{alg:pass}
\begin{algorithmic}[1]
\Function{BankPassing}{bank}
  \Statex \LeftComment{1} {\% Get current bank indices \%}
  \State {idx\_start, N\_local, N = \Call{BankScanning}{bank}}
  \State {idx\_end = idx\_start + N\_local}
  \Statex
  \Statex \LeftComment{1} {\% Get target indices based on N \%}
  \State {target\_start, target\_end = target\_idx(N)}
  \Statex
  \Statex \LeftComment{1}{\% Need more or less on each side? \%}
  \State {more\_left = idx\_start $<$ target\_start}
  \State {less\_left = idx\_start $>$ target\_start}
  \State {more\_right = idx\_end $>$ target\_end}
  \State {less\_right = idx\_end $<$ target\_end}
  \Statex
  \Statex \LeftComment{1} {\% Offside? \%}
  \State {offside\_left = idx\_end $\leq$ target\_start}
  \State {offside\_right = idx\_start $\geq$ target\_end}
  \Statex
  \Statex \LeftComment{1} {\% If offside, need to receive first \%}
  \If {offside\_left}
    \State {Receive from right}
    \State {less\_right = False}
  \EndIf
  \If {offside\_right}
    \State {Receive from left}
    \State {less\_left = False}
  \EndIf
  \Statex
  \Statex \LeftComment{1} {\% Non-blocking send \%}
  \If {more\_left}
    \State {Isend first (work\_start-idx\_start) particles to left}
  \EndIf
  \If {more\_right}
    \State {Isend last (work\_start-idx\_start) particles to right}
  \EndIf
  \Statex
  \Statex \LeftComment{1} {\% Receive (and wait if Isent) \%}
  \If {less\_left}
    \State {Receive from left}
  \EndIf
  \If {less\_right}
    \State {Receive from right}
  \EndIf
  \Statex
  \State \Return {bank\_final}
\EndFunction
\end{algorithmic}
\end{algorithm}

\begin{algorithm}[H] 
\singlespacing
\caption{Uniform Simple Sampling (SS)}
\label{alg:SS}
\begin{algorithmic}[1]
\Statex
\Function{PCT\_SS}{bank\_census, M}
  \State {idx\_start, N\_local, N = \Call{BankScanning}{bank\_census}}
  \Statex
  \Statex \LeftComment{1} {\% Count how many times particle is sampled \%}
  \State {count = \{0\}*N\_local}
  \For {i = 0:(M-1)}
    \State{idx = $\lfloor$random()*N$\rfloor$ - idx\_start}
    \If {0 $\leq$ idx $<$ N\_local}
      \State {count[idx] += 1}
    \EndIf
  \EndFor
  \Statex
  \Statex \LeftComment{1} {\% Set up the Sample Bank \%}
  \State {bank\_sample = \{\}}
  \For {i = 0:(N\_local-1)}
    \For {j = 1:count$[$i$]$}
      \State {particle = bank\_census[i]}
      \State {particle.w *= N/M}
      \State {bank\_sample.append(particle)}
    \EndFor
  \EndFor
  \Statex
  \State {bank\_final = \Call{BankPassing}{bank\_sample}}
  \State \Return {bank\_final}
\EndFunction
\end{algorithmic}
\end{algorithm}

\begin{algorithm}[H] 
\singlespacing
\caption{Uniform Splitting-Roulette (SR)}
\label{alg:SR}
\begin{algorithmic}[1]
\Statex
\Function{PCT\_SR}{bank\_census, M}
  \State {idx\_start, N\_local, N = \Call{BankScanning}{bank\_census}}
  \State {skip random number state by idx\_start}
  \State {n\_split = $\lfloor$M/N$\rfloor$}
  \State {p\_survive = M/N - n\_split}
  \Statex
  \Statex \LeftComment{1} {\% Split and then roulette each particle \%}
  \State {bank\_sample = \{\}}
  \For {i = 0:(N\_local-1)}
    \State {particle = bank\_census[i]}
    \State {particle.w *= N/M}
    \For {j = 1:n\_split}
      \State {bank\_sample.append(particle)}
    \EndFor
    \If {random() $<$ p\_survive}
      \State {bank\_sample.append(particle)}
    \EndIf
  \EndFor
  \Statex
  \State {skip random number state by (N-i\_start)}
  \State {bank\_final = \Call{BankPassing}{bank\_sample}}
  \State \Return {bank\_final}
\EndFunction
\end{algorithmic}
\end{algorithm}

\begin{algorithm}[H] 
\singlespacing
\caption{Uniform Particle Combing (CO)}
\label{alg:CO}
\begin{algorithmic}[1]
\Statex
\Function{PCT\_CO}{bank\_census, M}
  \State {idx\_start, N\_local, N = \Call{BankScanning}{bank\_census}}
  \State {idx\_end = idx\_start + N\_local}
  \State {tooth\_distance = N/M}
  \State {tooth\_offset = random()*tooth\_distance}
  \Statex
  \Statex \LeftComment{1} {\% First and last hitting teeth \%}
  \State {tooth\_start = $\lceil$(idx\_start-tooth\_offset)/tooth\_distance$\rceil$}
  \State {tooth\_end = $\lfloor$(idx\_end-tooth\_offset)/tooth\_distance$\rfloor$}
  \Statex
  \Statex \LeftComment{1} {\% Set the Sample Bank \%}
  \State {bank\_sample = \{\}}
  \For {i = tooth\_start:(tooth\_end+1)}
    \State {idx = $\lfloor$ i*tooth\_distance + tooth\_offset $\rfloor$ }
    \State {particle = bank\_census[idx - idx\_start]}
    \State {particle.w *= tooth\_distance}
    \State {bank\_sample.append(particle)}
  \EndFor
  \Statex
  \State {bank\_final = \Call{BankPassing}{bank\_sample}}
  \State \Return {bank\_final}
\EndFunction
\end{algorithmic}
\end{algorithm}

\begin{algorithm}[H]
\singlespacing
\caption{Uniform New Particle Combing (COX)}
\label{alg:COX}
\begin{algorithmic}[1]
\Statex
\Function{PCT\_COX}{bank\_census, M}
  \State {idx\_start, N\_local, N = \Call{BankScanning}{bank\_census}}
  \State {idx\_end = idx\_start + N\_local}
  \State {tooth\_distance = N/M}
  \Statex
  \Statex \LeftComment{1} {\% First and last possible hitting teeth \%}
  \State {tooth\_start = $\lfloor$idx\_start/tooth\_distance$\rfloor$}
  \State {tooth\_end = $\lceil$idx\_end/tooth\_distance$\rceil$ - 1}
  \State {skip random number state by tooth\_start}
  \Statex
  \Statex \LeftComment{1} {\% Sample and set the Sample Bank \%}
  \State {bank\_sample = \{\}}
  \For {i = tooth\_start:(tooth\_end+1)}
    \State {idx = $\lfloor$(random()+i)*tooth\_distance$\rfloor$ - idx\_start}
    \If {0 $\leq$ idx $<$ N\_local}
      \State {particle = bank\_census[idx]}
      \State {particle.w *= tooth\_distance}
      \State {bank\_sample.append(particle)}
    \EndIf
  \EndFor
  \Statex
  \State {skip random number state by (M-tooth\_start) }
  \State {bank\_final = \Call{BankPassing}{bank\_sample}}
  \State \Return {bank\_final}
\EndFunction
\end{algorithmic}
\end{algorithm}

\begin{algorithm}[H] 
\singlespacing
\caption{Uniform Duplicate-Discard (DD)}
\label{alg:DD}
\begin{algorithmic}[1]
\Statex
\Function{PCT\_DD}{bank\_census, M}
  \State {idx\_start, N\_local, N = \Call{BankScanning}{bank\_census}}
  \Statex
  \Statex \LeftComment{1} {\% Duplicate \%}
  \If {M $>$ N}
    \Statex \LeftComment{2} {\% Count how many times particle is sampled \%}
    \State {count = \{$\lfloor$M/N$\rfloor$\}*N\_local}
    \State {N\_sample = M - $\lfloor$M/N$\rfloor$*N }
    \For {i = 0:(N\_sample-1)}
      \State{idx = $\lfloor$random()*N$\rfloor$ - idx\_start}
      \If {0 $\leq$ idx $<$ N\_local}
        \State {count[idx] += 1}
      \EndIf
    \EndFor
    \Statex
    \Statex \LeftComment{2} {\% Set up the Sample Bank \%}
    \State {bank\_sample = \{\}}
    \For {i = 0:(N\_local-1)}
      \For {j = 1:count$[$i$]$}
        \State {particle = bank\_census[i]}
        \State {particle.w *= N/M}
        \State {bank\_sample.append(particle)}
      \EndFor
    \EndFor
  \Statex
  \Statex \LeftComment{1} {\% Discard \%}
  \Else
    \Statex \LeftComment{2} {\% Sample discarded particles and flag them \%}
    \State {N\_discard = N - M}
    \State {discard\_flag =  \{False\}*N}
    \For {i = 0:(N\_discard-1)}
      \Statex \LeftComment{3} {\% Rejection sampling \%}
      \While {True}
        \State{idx = $\lfloor$random()*N$\rfloor$}
        \If {discard\_flag[idx] = False}
          \State {discard\_flag[idx] = True}
          \State {break}
        \EndIf
      \EndWhile
    \EndFor
    \Statex
    \Statex \LeftComment{2} {\% Set up the Sample Bank \%}
    \State {bank\_sample = \{\}}
    \For {i = idx\_start:idx\_start+N\_local}
      \If {discard\_flag[idx] = False}
        \State {idx\_local = i - idx\_start}
        \State {particle = bank\_census[idx\_local]}
        \State {particle.w *= N/M}
        \State {bank\_sample.append(particle)}
      \EndIf
    \EndFor
  \EndIf
  \Statex
  \State {bank\_final = \Call{BankPassing}{bank\_sample}}
  \State \Return {bank\_final}
\EndFunction
\end{algorithmic}
\end{algorithm}

\end{document}